\documentclass[12pt]{article}

\usepackage{natbib}
\usepackage{multibib} 
\usepackage{anysize}
\usepackage{setspace}
\usepackage[T1]{fontenc}
\usepackage{lmodern}
\usepackage{hyperref}
\usepackage{graphicx}	
\usepackage{tabularx}	
\usepackage{subcaption} 
\usepackage{amssymb,amsmath,amsthm}
\usepackage{booktabs} 
\usepackage{cleveref}		
\crefname{figure}{figure}{figures}
\usepackage{chngcntr} 	
\usepackage[usenames, dvipsnames]{color}

\usepackage{array}
\newcolumntype{L}[1]{>{\raggedright\let\newline\\\arraybackslash\hspace{0pt}}m{#1}}
\newcolumntype{C}[1]{>{\centering\let\newline\\\arraybackslash\hspace{0pt}}m{#1}}
\newcolumntype{R}[1]{>{\raggedleft\let\newline\\\arraybackslash\hspace{0pt}}m{#1}}

\title{Externalities in Knowledge Production:\\ Evidence from a Randomized Field Experiment\thanks{We are grateful to Chris Forman, Willa Friedman, Shane Greenstein, David Hugh-Jones, Giovanni Mastrobuoni, Ignacio Monzon, Juan S. Morales, Abhishek Nagaraj, Stephan Seiler, Ananya Sen, and Michael Zhang for valuable comments.}}

\author{
Marit Hinnosaar%
\thanks{Collegio Carlo Alberto and CEPR, \href{mailto:marit.hinnosaar@gmail.com}{\url{marit.hinnosaar@gmail.com}}.%
}
\and
Toomas Hinnosaar%
\thanks{Collegio Carlo Alberto, \href{mailto:toomas@hinnosaar.net}{\url{toomas@hinnosaar.net}}. %
}
\and
Michael Kummer%
\thanks{Georgia Institute of Technology \& University of East Anglia, \href{mailto:michael.kummer@econ.gatech.edu}{\url{michael.kummer@econ.gatech.edu}}.%
}
\and
Olga Slivko%
\thanks{Centre for European Economic Research (ZEW), \href{mailto:slivko@zew.de}{\url{slivko@zew.de}}.%
}
}

\begin{document}
\maketitle
\begin{abstract}
Do contributions to online content platforms induce a feedback loop of ever more user-generated content or will they discourage future contributions? To assess this, we use a randomized field experiment which added content to some pages in Wikipedia while leaving similar pages unchanged. We find that adding content has a negligible impact on the subsequent long-run growth of content. Our results have implications for information seeding and incentivizing contributions, implying that additional content does not generate sizable externalities, neither by inspiring nor by discouraging future contributions.
\end{abstract}


\section{Introduction}

Knowledge is a key input to many economic activities and a driver of economic growth \citep{romer1990endogenous,grossman1993innovation,jones1995r}. An increasing share of knowledge is created in the form of user-generated content: consumer feedback systems, discussion boards, Q\&A sites, open-source software, social networks, and online information repositories, such as Wikipedia. Understanding the drivers of contributions to user-generated content has been an important question in economics and management for the past two decades \citep{lerner-tirole-2002}.

In this paper, we investigate whether there are positive or negative externalities in user-generated content production.
Understanding and quantifying such externalities has both policy and managerial implications. If content generation has positive externalities on future content generation, then information seeding\footnote{\cite{nagaraj2017information} describes how such policies have been used by Wikipedia (seeding articles on more than 30,000 US cities from US Census Bureau data), OpenStreetMap (US Census maps), and Reddit (fake user accounts).} and paid contributions may have a high return on investment in terms of added stimulated growth \citep{aaltonen-seiler-2016}.
On the other hand, if content generation has negative externalities, then such policies may backfire and be not only ineffective but lead to worse eventual outcomes \citep{nagaraj2017information}.

Due to the reflection problem \citep{manski-1993}, externalities in content generation are difficult to identify. An externality occurs when a contribution by a user motivates other users to contribute (positive externality) or prevents further contributions (negative externality). Yet, the correlation in their contributions does not necessarily attest to an externality. A positive correlation may arise when users contribute because they were both exposed to the same external shock, such as a news article or a research finding. Similarly, a non-causal negative correlation may be caused by processes with periodic updates, such as elections or periodically updated statistics. To identify the causal effect, shocks to content growth and contributions must be independent over time. Without randomization provided by an experiment, the assumption is unlikely to be satisfied.

We estimate the causal impact of additional content on subsequent contributions using a randomized field experiment in Wikipedia. Randomization ensures that the addition of content is exogenous in terms of future content generation. As argued by the literature analyzing social interactions (including \cite{manski-1993} and \cite{aaltonen-seiler-2016}), randomized experiments are the best way to cleanly identify causal relationships in such interactions.

The exogenous variation in our data is generated by a randomized field experiment, which was conducted in 2014.\footnote{A comprehensive description of the experiment is provided in \cite{hinnosaar2017wikipedia}, who studied the impact of this treatment on real-world outcomes.} The experiment added relevant content to randomly chosen Wikipedia pages while leaving similar pages unchanged. The treatment added about two paragraphs (approximately 2,000 characters) and one picture to each page in the treatment group. The pages were about mid-sized Spanish cities in different language editions of Wikipedia.

We use a dataset of Wikipedia editing histories which includes all versions of Wikipedia pages in the treatment and the control groups. Our dataset and the experimental setting allow us to analyze both short-term and long-term effects, up to four years after the experiment. Our main variable of interest is the outcome of content production---page length. To study the impact on the quality of content, we also analyze various other measures of editing activity, including the number of unique editors, number of edits, and the amount of content added and deleted.

Our main finding is that the additional content has a negligible impact on the subsequent long-run growth. The pages which were improved by adding about 2,000 characters of content, four years later are still longer by about the same amount. 
However, we do find some evidence that in the short-run the editing activity increased. In the first two years after the experiment, the treatment increased the number of Wikipedia users editing the treated pages and increased the number of edits. The increase in the number of users and edits was short-lived; there was no impact in the third and fourth year post-experiment. Moreover, the amount of content these users added was small and their edits were mostly limited to directly modifying the text added by the treatment. Our findings are robust to a set of alternative specifications, including alternative covariates and both cross-sectional and panel data frameworks. The experimental variation allows interpreting our findings as causal effects. 

These findings have a clear policy and managerial implication---investments in information seeding and additional content contributions have a negligible cumulative effect on growth. Therefore information seeding and incentivizing contributions are a matter of direct cost-benefit analysis: they pay off if and only if the costs of creating the content are lower than the value of the new content. The additional costs or benefits via externalities that discourage or inspire future contributions are negligible.

Our paper contributes to the literature that studies externalities in user-generated content production. The closest to our work are \cite{aaltonen-seiler-2016} and \cite{nagaraj2017information}. \cite{aaltonen-seiler-2016} used detailed observational data from Wikipedia.  \cite{nagaraj2017information} used a natural experiment on OpenStreetMap, a Wikipedia-style digital map-making community, which started out with better seeding information in some regions compared to others for quasi-random reasons. The two papers arrive at contradicting conclusions, which warrant further investigation regarding this issue. Our paper is the first to study the question using a randomized field experiment, which allows causal identification of the underlying externalities. 

More generally, the paper belongs to the literature that analyzes what drives contributions to user-generated content.\footnote{Other studies on Wikipedia have analyzed biases in Wikipedia's content \citep{greenstein_is_2012,greenstein_experts_2017,hinnosaar2018gender} and the impact of Wikipedia on market outcomes \citep{xu_impact_2013,hinnosaar2017wikipedia} and science \citep{thompson2018science}.
}
The topics addressed in the prior related literature include the role of personal gain \citep{shah_motivation_2006}, group size \citep{zhang_group_2011}, networks \citep{fershtman_direct_2011, ransbotham_network_2012}, spillovers \citep{kummer2014spillovers}, symbolic awards \citep{gallus_fostering_2017},  performance feedback \citep{huang_motivating_2018}, monetary rewards vs social motives \citep{sun_motivation_2017}, contributor diversity \citep{ren_impact_2015}, and economic conditions, such as unemployment \citep{kummer2015downturn} and migration \citep{slivko2018brain}.
More closely related to our work is \cite{kane_content_2016} who study the relationship between contributions and consumption of content. Using a large observational dataset from Wikipedia they document that more content increases contributions but the effect decreases over time. 
Our paper extends the literature by using variation from a randomized field experiment to measure the impact of additional content on future content generation. It confirms the finding by \cite{kane_content_2016} that the impact decreases over time, moreover, it finds that the long-term effect is negligible.  

The structure of the paper is as follows. In the next section, we describe the experiment. \Cref{S:Data} describes the data. \Cref{S:Results} describes the empirical strategy and presents the results of the impact of treatment on the subsequent page length and on editing activity: the number of editors, the number of edits, and how much they edited. \Cref{S:Discussion} discusses the connection between our findings and previous literature and limitations in our analysis. \Cref{S:Conclusions} concludes.

\section{Experiment} \label{S:Experiment}

The field experiment added content (text and photos) to randomly chosen Wikipedia pages.
The sample consisted of 240 Wikipedia pages. Specifically, it consisted of the pages of 60 Spanish cities in the French, German, Italian, and Dutch editions of Wikipedia. The cities were all medium sized, excluding the largest like Madrid and Barcelona, and excluding also the smaller cities. The Wikipedia pages in these languages were relatively short---up to 24,000 characters in each of these four languages.

Each city and each language edition of Wikipedia was treated equally. For each city, its page was assigned to the treatment group in two randomly chosen languages. In each language edition of Wikipedia, 30 randomly chosen city pages were assigned to the treatment group. Specifically, to obtain balance in the treatment and control groups, the randomization was stratified.\footnote{For further details of the randomization, see \cite{hinnosaar2017wikipedia}.}
The 60 cities were divided into ten equal-sized groups. Within each group, each city was randomly assigned to one of six treatment arms. The six treatment arms were as follows: treat the city page in one of the six possible language pairs (French \& German; French \& Italian; French \& Dutch; German \& Italian; German \& Dutch; Italian \& Dutch). This resulted in a design where the number of pages which were treated equaled the number of those that remained in the control group.

The Wikipedia pages were treated mid-August, 2014. The treatment added about 2,000 characters of text and photos to each page in the treatment group. The added text and photos were mostly obtained from the corresponding Spanish and English language Wikipedia pages. Because all the pages were about Spanish cities, the Spanish Wikipedia typically contained more information than the other language versions. The English language version of the page, typically, was also more detailed than in the languages in the experiment. Hence, there was information available in Spanish and English pages that was missing from the other language editions of Wikipedia. The treatment translated that text and added it to the corresponding pages in the treatment group.

The treatment of the pages in Dutch Wikipedia was not successful. While in French, German, and Italian Wikipedia, the added text and photos survived well over time, all the additions to Dutch Wikipedia were deleted within 24 hours (by a single editor). Wikipedia allows anyone to edit. It also means that anyone can delete or undo the latest changes by reverting to a previous version of the page. This happened in the Dutch version of Wikipedia, where 24 hours after the treatment, all the pages looked as if they had never been treated. Therefore, we exclude Dutch pages from our analysis and restrict attention to the 180 pages in French, German, and Italian. Note that our main results do not change if Dutch pages are included in the analysis.

\section{Data} \label{S:Data}

Our main sources of data are the Wikipedia editing histories in the treated languages. An editing history contains the full text of each revision\footnote{A revision (or an edit) is a version of a Wikipedia article saved at a specific moment of time by a specific user. All revisions with the corresponding metadata, including full text, user, and timestamp, are preserved by Wikipedia and publicly available.} of each page starting from the creation of the page until the beginning of September 2018. Our sample consists of the 180 pages in the experiment, which are the pages of 60 cities in French, German, and Italian Wikipedia.
In the following subsections, we describe the construction of the dataset and variables used in the analysis.

\subsection{Page length}

Our main outcome variable is the page length after the experiment. We measure page length in characters, including spaces and wiki markup commands.

\begin{figure}[h]
\begin{center}
    \includegraphics[width=0.7\linewidth]{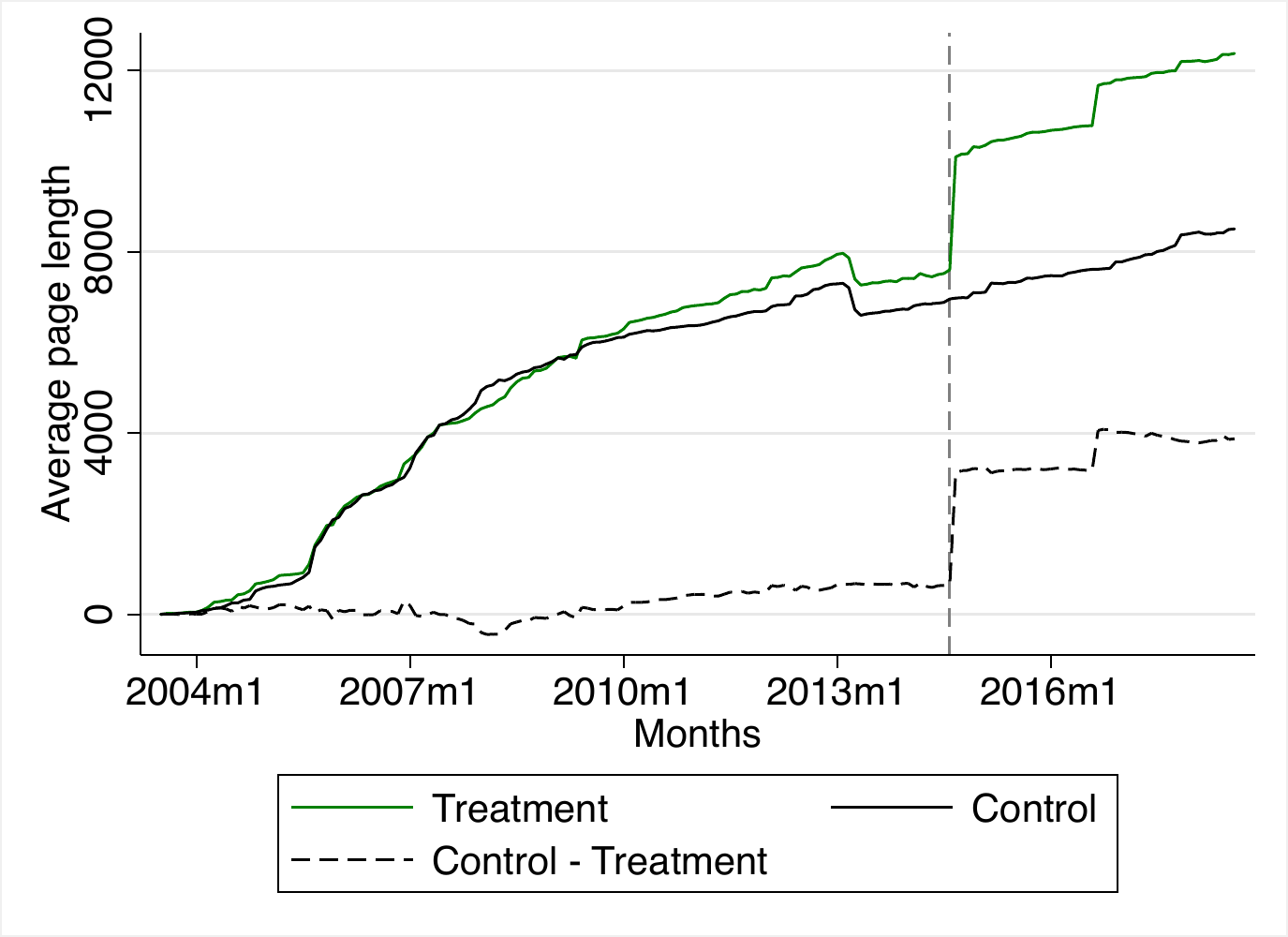}
    \caption{Average page length in the treatment and control groups}
    \label{F:length}
\end{center}
\footnotesize{
    Notes: The number of observations is 90 in the control and 90 in the treatment groups. The experiment month (August 2014) is marked by the dashed vertical line.
}
\end{figure}

\Cref{F:length} presents average page length in the treatment and control groups. Until the experiment in August 2014, the average page length in the control and treatment groups was rather similar.\footnote{The drop in both the treatment and control groups in early 2013 comes from technical changes in Wikipedia: Addbot removed about 2,000 characters from each page with an explanation similar to ``Migrating 77 interwiki links, now provided by Wikidata''.}  The experiment added significant length to the pages in the treatment group. After the experiment, the difference has been relatively stable. An exception is a sharp increase in the mean of the treatment group in August 2016. This jump comes from the efforts of a single editor who worked hard to improve one page in French Wikipedia---the page of the city of Cordoba.\footnote{By August 2016, the page of Cordoba in French Wikipedia was relatively mature, with length 19,426 characters (at the time 93\% of the pages in our sample were shorter than that). During August 2016 this user increased the page length to 100,702 characters which is almost twice the length of the longest page at the time (57,076 characters). Our conclusions do not change if we exclude this page. }
\Cref{A:AdditionalFiguresTables} presents the same figure first, without French Cordoba (\cref{F:Length_wo_FrenchCordoba}) and second, with the logarithm of page length (\cref{F:LogLength}), both of which show no evidence of an increase in the treatment group average in 2016.

\begin{figure}[!ht]
\begin{center}
  \begin{subfigure}[b]{0.49\textwidth}
	\includegraphics[width=\textwidth]{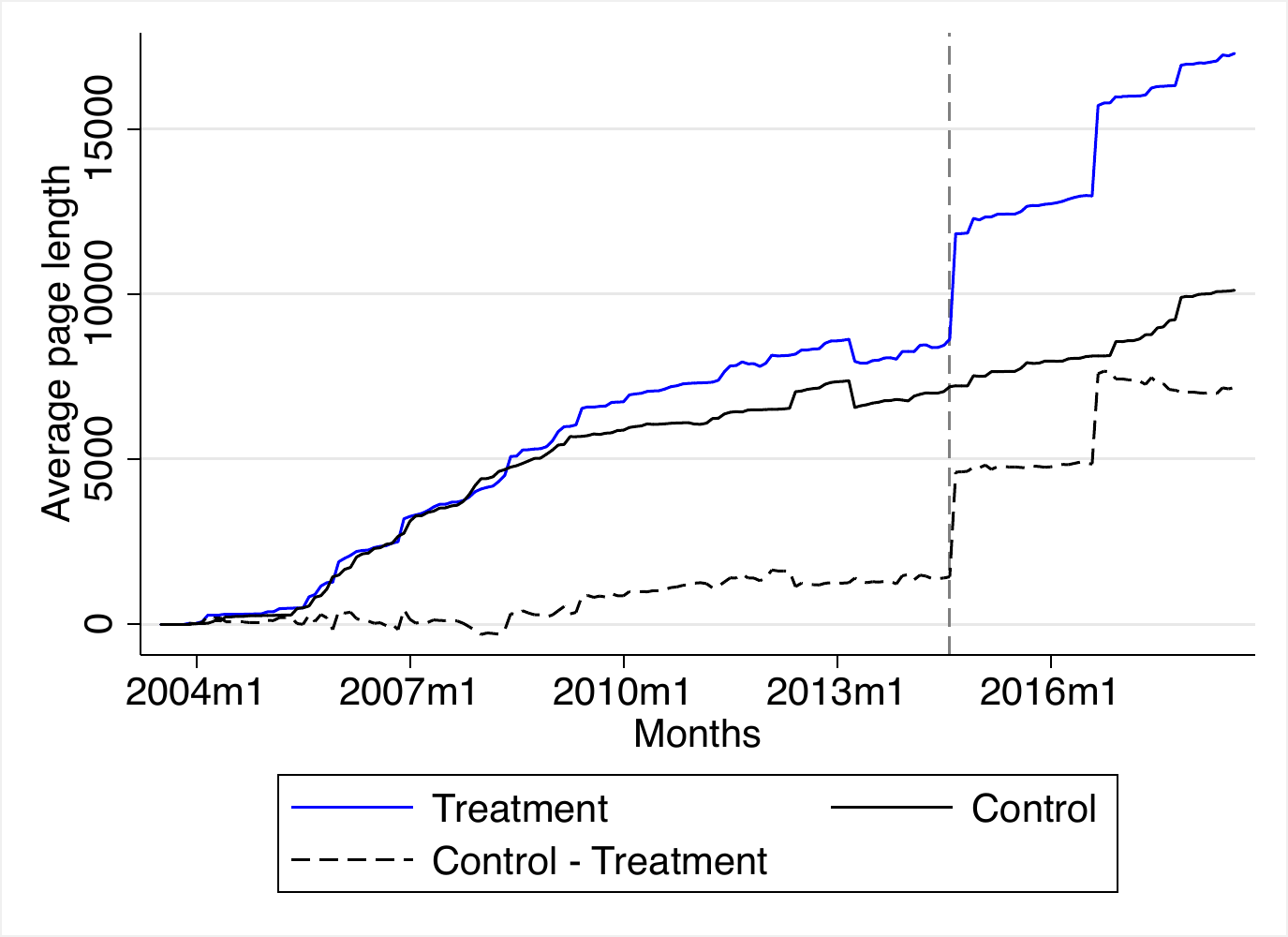}
  	\caption{French}
  	\label{F:Length_fr}
  \end{subfigure}
  \begin{subfigure}[b]{0.49\textwidth}
	\includegraphics[width=\textwidth]{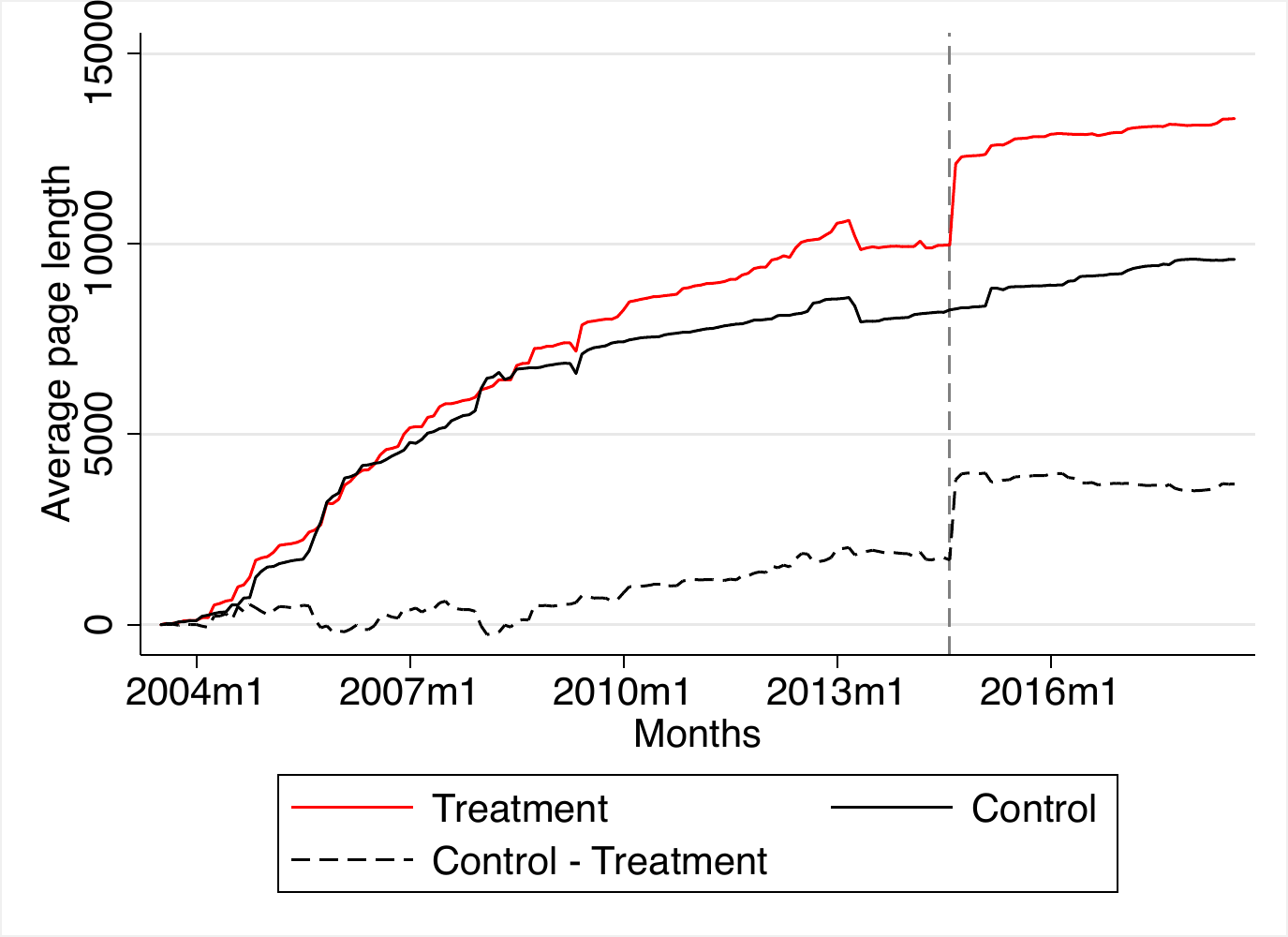}
  	\caption{German}
  	\label{F:Length_de}
  \end{subfigure}
  \begin{subfigure}[b]{0.49\textwidth}
	\includegraphics[width=\textwidth]{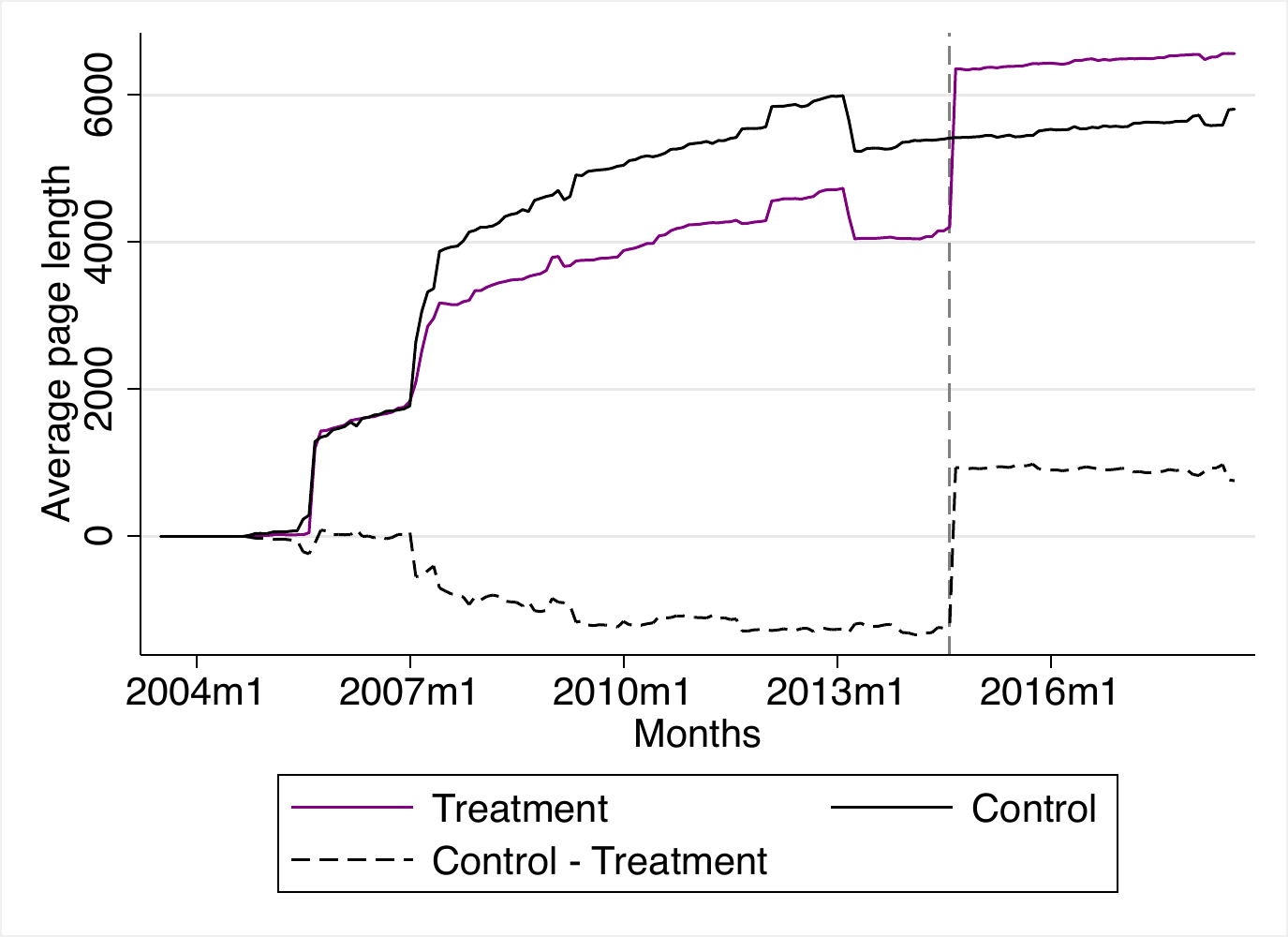}
  	\caption{Italian}
  	\label{F:Length_it}
  \end{subfigure}
  \begin{subfigure}[b]{0.49\textwidth}
	\includegraphics[width=\textwidth]{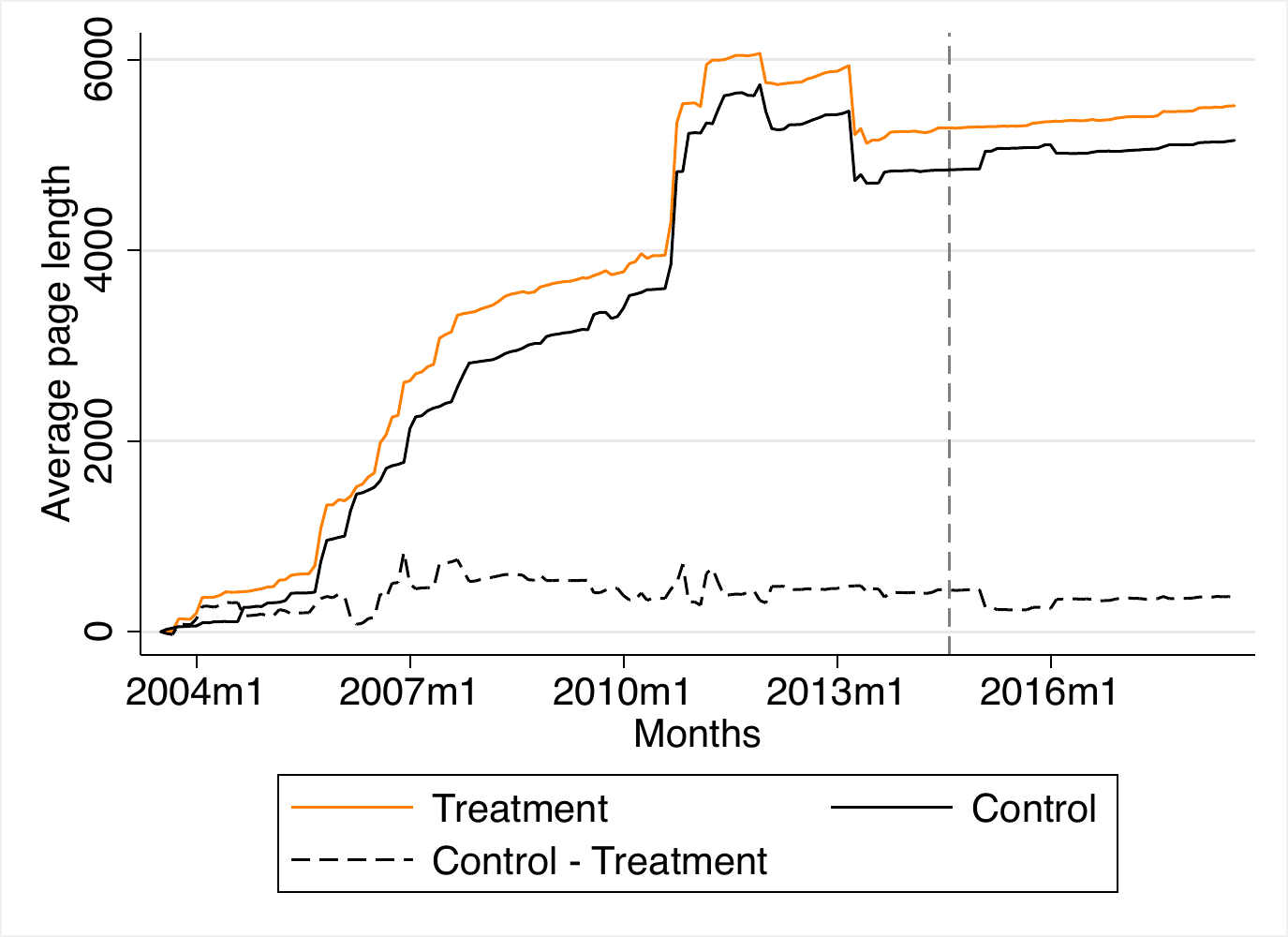}
  	\caption{Placebo: Dutch}
  	\label{F:Length_nl}
  \end{subfigure}
  \caption{Average page length in the treatment and control groups, by language}
  \label{F:length_lang}
\end{center}
\footnotesize{
    Notes: On each figure, the number of observations is 30 in the control and 30 in the treatment groups.
    The experiment month (August 2014) is marked by the dashed vertical line.
}
\end{figure}

Similar dynamics can be seen when looking at the changes separately by language (\cref{F:Length_fr,F:Length_de,F:Length_it}). As expected, the placebo test with the Dutch pages (\cref{F:Length_nl}) shows that the assignment to treatment group had no impact. Page length is one possible output measure of knowledge production in Wikipedia. Similar dynamics as on \cref{F:length}  can also be seen on \cref{F:LengthImagesPlainText} in \cref{A:AdditionalFiguresTables} which presents alternative measures of content: images and plain text (that is, html elements removed from the parsed text).

To make the treatment and the control groups comparable we subtracted the length of text added by the treatment from the length of pages in the treatment group.
\Cref{T:SummaryStatistics} presents summary statistics of page length after the experiment.

\begin{table}[ht!]
\renewcommand{\tabcolsep}{10pt}
\begin{center}
\caption{Summary statistics of page length and editing activity post-treatment}
\label{T:SummaryStatistics}
\begin{tabular}{ l cccc cccc }
\hline
 Variable & Mean & SE & Min & 10th & 50th & 90th & Max  & Obs \\ \hline \multicolumn{9}{c}{Panel A: Page length minus the length of text added by treatment, post-treatment} \\
1st year        &     8012&     6447&     1260&     1696&     6588&    16168&    55378&      180\\
2nd year        &     8640&     9356&      646&     1663&     6906&    16302&    98047&      180\\
3rd year        &     9037&     9709&      627&     1744&     7358&    17250&   101428&      180\\
4th year        &     9463&    10141&      611&     1688&     7554&    18758&   102688&      180\\
 \hline \multicolumn{9}{c}{Panel B: Average monthly number of unique users, post-treatment} \\ 
1st year        &     0.37&     0.33&     0.00&     0.00&     0.33&     0.75&     1.67&      180\\
2nd year        &     0.32&     0.35&     0.00&     0.00&     0.17&     0.71&     2.58&      180\\
3rd year        &     0.30&     0.29&     0.00&     0.00&     0.25&     0.67&     1.58&      180\\
4th year        &     0.29&     0.28&     0.00&     0.00&     0.17&     0.71&     1.25&      180\\
 \hline \multicolumn{9}{c}{Panel C: Average monthly number of edits, post-treatment} \\ 
1st year        &     0.39&     0.35&     0.00&     0.00&     0.33&     0.75&     1.83&      180\\
2nd year        &     0.34&     0.41&     0.00&     0.00&     0.21&     0.75&     3.67&      180\\
3rd year        &     0.32&     0.33&     0.00&     0.00&     0.25&     0.75&     2.00&      180\\
4th year        &     0.30&     0.30&     0.00&     0.00&     0.17&     0.75&     1.33&      180\\
 \hline \multicolumn{9}{c}{Panel D: Average monthly edit distance, post-treatment} \\ 
1st year        &   119.85&   205.39&     0.00&     0.00&    27.96&   358.21&  1520.83&      180\\
2nd year        &   108.32&   678.72&     0.00&     0.00&    13.71&   187.29&  9027.83&      180\\
3rd year        &    65.12&   145.85&     0.00&     0.00&    12.92&   183.29&  1273.92&      180\\
4th year        &   112.12&   500.08&     0.00&     0.00&    11.12&   167.58&  5800.50&      180\\

\hline
\end{tabular}
\end{center}
\footnotesize{
Notes: A unit of observation is a page (180 pages).
Panel A reports summary statistics of page length (minus the length of text added by treatment) at the end of the 1st, 2nd, 3rd, and 4th year post-treatment. Summary statistics of post-treatment average monthly number of unique users are in panel B, average monthly number of edits in panel C, and average monthly edit distance in panel D. Page length and edit distance are measured in the number of characters. 
}
\end{table}

\subsection{Measures of editing activity}

To construct the measures of editing activity, we start with 30,601 edits (revisions) from 180 Wikipedia pages. This includes all the edits except those generated as part of the treatment in the experiment. Following \cite{aaltonen-seiler-2016}, we restrict the sample of edits in the following ways. First, we exclude edits by bots (about 30\% of edits), these are non-human user accounts that generate automated edits. Specifically, we define bots as users whose username occurs in the list of bots (in the English, French, German or Dutch Wikipedias) or whose username includes ``bot''. Second, we exclude reverts, which are edits that restore any previous version of the same page (about 7\% of remaining edits). Third, we exclude vandalism (about 0.8\% of remaining edits). We use the following criteria to classify an edit as vandalism: (a) an edit that only deletes text from the previous revision, and (b) the revision immediately after vandalism reverts the article back to a past revision. Then we are left with 19,586 productive edits generated by human users.

To analyze the impact of treatment on editing activity we construct three types of monthly measures that characterize how many people edited the pages, how many times they edited, and how much they edited. The first measure is the number of unique users editing a page per month. We define a unique user by the username for registered users and by IP address for anonymous users. The second measure is the number of edits per month. To avoid double-counting of micro-edits,\footnote{Many Wikipedia editors save many revisions to the same page in a short period of time (because they save a revision after every small change while editing).} we first aggregate edits to day-user-page level, and then sum these up to month-page level.
The third measure is edit distance, which we define as the number of characters an edit added plus the number of characters it deleted compared to the previous version of the page.\footnote{Note that our edit distance measure differs from the Levenshtein edit distance by giving weight two to substitutions instead of weight one. For each edit, we calculate the edit distance using php FineDiff class at the granularity level of a character.} We aggregate the edit distance measure to monthly level.

Panels B to D in \cref{T:SummaryStatistics} present summary statistics of editing activity.
\Cref{F:EditingActivity} in \cref{A:AdditionalFiguresTables}  describes the average editing activity in the treatment and control groups over time. \Cref{T:BalanceTest} in \cref{A:AdditionalFiguresTables} presents the comparison of pre-treatment page length and editing activity in the treatment group versus the control group. The table shows that there was no significant difference between the two groups before the treatment.

In addition to the aggregate measures of editing activity, we separate edits that directly modify the treatment text and those that modify other parts of the page. We classify edits into these two categories using a method similar to \cite{hinnosaar2017wikipedia}. 
For each page in the treatment group, we use the diff algorithm between the revision before and after the treatment to determine treatment text---the exact text added by the treatment.
For each revision post-treatment, using the diff algorithm between the treatment text and this revision, we check whether the revision deleted any part of the treatment text. If the revision didn't delete anything from the treatment text, we classify the revision as one that edited other parts of the page.

\section{Results}\label{S:Results}
\subsection{The impact on page length} \label{S:ImpactOnOutcomes}

\paragraph{Main empirical strategy.}
Because we are able to use experimental variation, we focus on cross-sectional estimation.
Our empirical strategy compares page lengths after the experiment in the treatment and control groups.
To make the pages comparable, we subtract the length of text added by the treatment from the length of pages in the treatment group.
Hence, the estimates should be interpreted as the effect of treatment on page length after removing the mechanical increase created by the treatment.
We estimate the following regression:
\begin{equation}\label{E:CrossSectionRegression}
Y_i = \beta_0 + \beta_1 TreatmentGroup_i + X_i +\varepsilon_i
\end{equation}
where the outcome variable is the logarithm of page length (from which the length of the treatment text has been subtracted) of page $i$ in the years post-experiment. The coefficient of interest is $\beta_1$ on $TreatmentGroup_i$, which is an indicator variable that takes value one if the page was assigned to the treatment group and zero if it was assigned to the control group. Covariates, $X_i$, include language and city fixed effects and logarithm of pre-treatment page length.

\paragraph{Main results.} \Cref{T:RegrLogLengthCrossSection} presents the main results of the impact of treatment on page length up to four years after the experiment. Specifically, the outcome variables are the logarithm of page length (minus treatment text) at the end of the 1st year post-treatment (panel A), 2nd, 3rd, and the 4th year post-treatment (panels B to D). Columns 1--4, starting from no controls, successively add all the controls. The coefficients are stable across the specifications while adding covariates decreases standard errors. 

\begin{table}[h!]
\begin{center}
\caption{The effect of treatment on page length}
\label{T:RegrLogLengthCrossSection}
\renewcommand{\tabcolsep}{2pt}
\begin{tabular}{l cccc@{\hskip 0.2in} cccc}
\hline
  & \multicolumn{8}{c}{ Log. page length (minus treatment text)} \\ & \multicolumn{4}{c}{Panel A: 1st year post-treatment} & \multicolumn{4}{c}{Panel B: 2nd year post-treatment} \\  & (1) & (2) & (3) & (4) & (1) & (2) & (3) & (4) \\ \cmidrule(lr){2-5} \cmidrule(lr){6-9} 
Treatment group     &       0.068   &       0.069   &       0.045   &       0.065*  &       0.071   &       0.075   &       0.047   &       0.071*  \\
                    &     (0.118)   &     (0.088)   &     (0.036)   &     (0.035)   &     (0.125)   &     (0.091)   &     (0.043)   &     (0.042)   \\
Log. length before  &               &               &       0.927***&       0.851***&               &               &       0.962***&       0.853***\\
treatment           &               &               &     (0.022)   &     (0.034)   &               &               &     (0.027)   &     (0.041)   \\
Language FE         &          No   &         Yes   &          No   &         Yes   &          No   &         Yes   &          No   &         Yes   \\
City FE             &          No   &         Yes   &          No   &         Yes   &          No   &         Yes   &          No   &         Yes   \\
Mean dep. var.      &       8.699   &       8.699   &       8.699   &       8.699   &       8.723   &       8.723   &       8.723   &       8.723   \\
SD dep. var.        &       0.792   &       0.792   &       0.792   &       0.792   &       0.836   &       0.836   &       0.836   &       0.836   \\
Adj. R-squared      &      -0.004   &       0.510   &       0.909   &       0.922   &      -0.004   &       0.526   &       0.880   &       0.898   \\
Observations        &         180   &         180   &         180   &         180   &         180   &         180   &         180   &         180   \\
\hline & \multicolumn{8}{c}{ Log. page length (minus treatment text)} \\ & \multicolumn{4}{c}{Panel C: 3rd year  post-treatment} & \multicolumn{4}{c}{Panel D: 4th year  post-treatment} \\  & (1) & (2) & (3) & (4) & (1) & (2) & (3) & (4) \\ \cmidrule(lr){2-5} \cmidrule(lr){6-9} 
Treatment group     &       0.059   &       0.074   &       0.035   &       0.069   &       0.040   &       0.048   &       0.017   &       0.045   \\
                    &     (0.127)   &     (0.092)   &     (0.048)   &     (0.045)   &     (0.130)   &     (0.092)   &     (0.056)   &     (0.049)   \\
Log. length before  &               &               &       0.963***&       0.857***&               &               &       0.963***&       0.830***\\
treatment           &               &               &     (0.029)   &     (0.044)   &               &               &     (0.034)   &     (0.048)   \\
Language FE         &          No   &         Yes   &          No   &         Yes   &          No   &         Yes   &          No   &         Yes   \\
City FE             &          No   &         Yes   &          No   &         Yes   &          No   &         Yes   &          No   &         Yes   \\
Mean dep. var.      &       8.762   &       8.762   &       8.762   &       8.762   &       8.795   &       8.795   &       8.795   &       8.795   \\
SD dep. var.        &       0.847   &       0.847   &       0.847   &       0.847   &       0.868   &       0.868   &       0.868   &       0.868   \\
Adj. R-squared      &      -0.004   &       0.524   &       0.858   &       0.889   &      -0.005   &       0.547   &       0.816   &       0.873   \\
Observations        &         180   &         180   &         180   &         180   &         180   &         180   &         180   &         180   \\

\hline
\end{tabular}
\end{center}
\footnotesize{
Notes:
Each column presents estimates from a separate cross-section regression of 180 Wikipedia pages. The dependent variable is the logarithm of page length (minus the length of text added by the treatment) in the 1st year post-treatment (panel A), 2nd (panel B), 3rd (panel C), and the 4th year post-treatment (panel D).
Standard errors are reported in parentheses. *** Indicates significance at 1 percent level, ** at 5 percent level, * at 10 percent level.
}
\end{table}

The results in \cref{T:RegrLogLengthCrossSection} indicate that, once all the covariates have been included, pages in the treatment group compared to those in the control in the first years post-treatment are about 7\% longer. By the fourth year, the treatment effect is not statistically significant, and the coefficient is smaller in magnitude. In all the years, the coefficient  of the impact of treatment is less than 9\% of the standard deviation (reported in  \cref{T:RegrLogLengthCrossSection}).  

\begin{figure}[!ht]
\begin{center}
\includegraphics[width=0.7\textwidth]{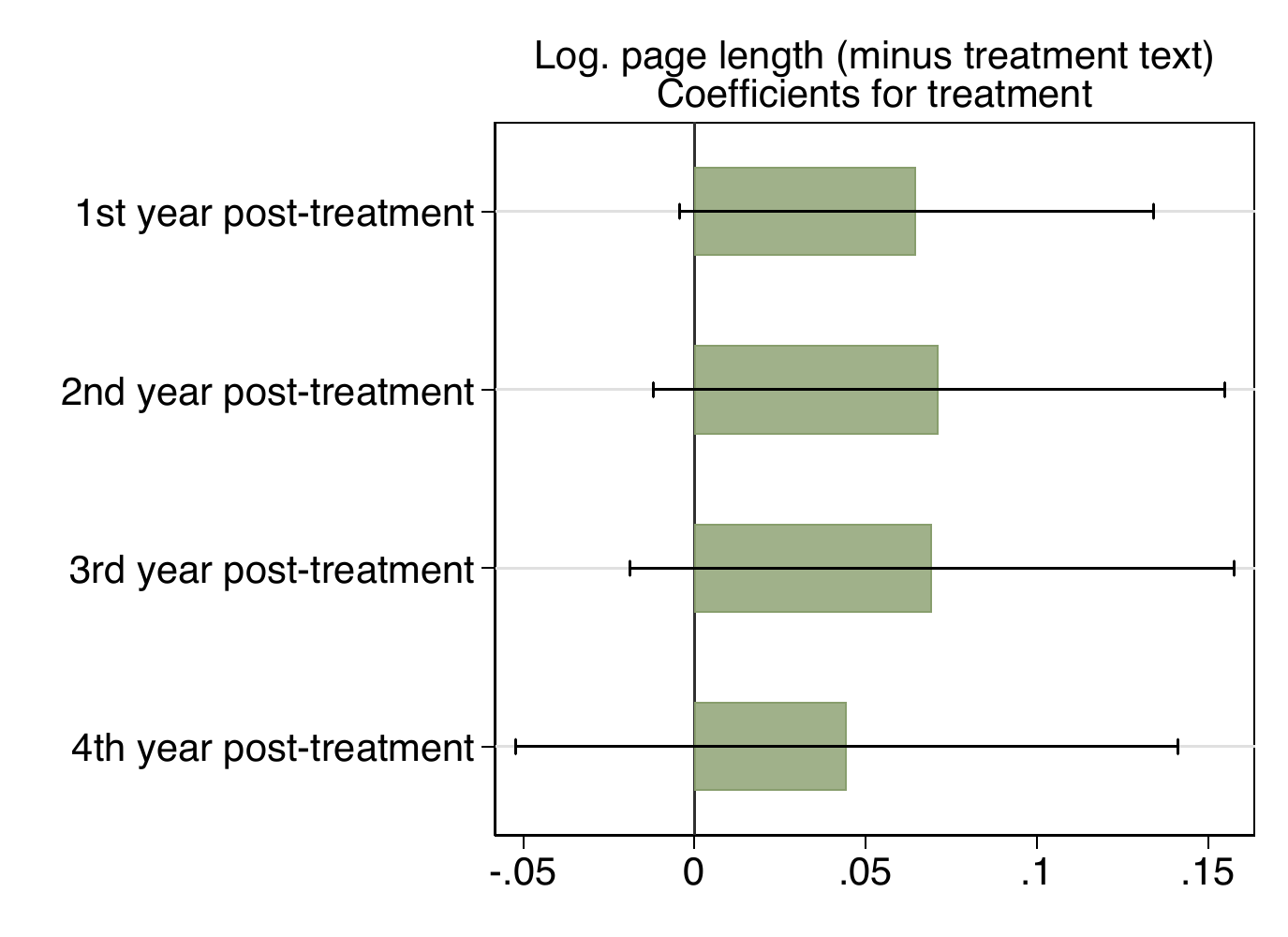}
\caption{The impact of treatment on page length}
\label{F:RegrLogLengthCrossSection}
\end{center}
\footnotesize{
Notes: This figure presents point estimates (bars) and 95\% confidence intervals (lines) of coefficients from  four regressions in column 4 of \cref{T:RegrLogLengthCrossSection}. The coefficients describe the impact of treatment on the logarithm of page length (from which the length of text added by the treatment has been subtracted) at the end of the 1st, 2nd, 3rd, and 4th year post-treatment. Unit of observation is a page (180 pages). For further details see \cref{T:RegrLogLengthCrossSection}.
}
\end{figure}

The results in \cref{T:RegrLogLengthCrossSection} suggest that by the end of the fourth year the impact of treatment on page length is negligible. 
After excluding the length of text added by the treatment, using the bounds of the 95\% confidence interval implied by estimates in column 4 panel D, we are able to reject that by the end of the 4th year after the experiment, pages in the treatment group are more than 14.1\% longer or more than 5.1\% shorter than those in the control group.\footnote{The upper bound: $0.045 + 1.96\times 0.049 = 0.141$; and the lower bound: $0.045-1.96\times 0.049 = -0.051$.} 
The estimation results are summarized graphically on \cref{F:RegrLogLengthCrossSection}.

\paragraph{Robustness and heterogeneity.}
Our analysis of robustness and heterogeneity is provided in online \cref{A:AdditionalFiguresTables}.
We address 
robustness to inclusion of alternative covariates (\cref{T:RegrRobustLogLengthAlternControls})
and heterogeneity across languages (\cref{T:RegrRobustLogLengthByLanguage}) and across page characteristics (\cref{T:RegrHeterLogLengthPageAge}). In brief, we do not find any evidence that the treatment had any significant long term effect on page length.

First, we assess whether the results are sensitive to inclusion of alternative covariates. \Cref{T:RegrRobustLogLengthAlternControls} presents estimates from the same regressions as in \cref{T:RegrLogLengthCrossSection}, but instead of city and language fixed effects, it includes stratification group dummies (see \cref{S:Experiment}). Results are identical to those without any fixed effects in \cref{T:RegrLogLengthCrossSection} (columns 1 and 3 in \cref{T:RegrLogLengthCrossSection}).

\Cref{T:RegrRobustLogLengthByLanguage} re-estimates the regressions separately for each language. There is a small significant effect of treatment on page length only in French Wikipedia and only in the first two years post-treatment. There is no significant effect in other languages.  \Cref{T:RegrHeterLogLengthPageAge} re-estimates the regressions while dividing the sample by page age. While the estimates are insignificant, the magnitude of the coefficients in the first years post-treatment is larger in the case of younger pages.

\subsection{The impact on the number of users and edits}\label{S:ImpactOnUsersEdits}

\paragraph{Main results. } 
\Cref{T:RegrUsersCrossSection} presents the main results of the impact of treatment on the subsequent number of users and edits up to four years after the experiment. Each column presents estimates from similar cross-sectional regressions as in \cref{S:ImpactOnOutcomes}, where the outcome variables measure editing activity in the 1st, 2nd, 3rd, and the 4th year after the experiment. The outcome variables are the average number of users (people editing the page) per month (panel A) and the average number of edits per month (panel B). In our preferred specification in \cref{T:RegrUsersCrossSection}, covariates include city and language fixed effects. Additional covariates are included in the robustness analysis in online \cref{A:AdditionalFiguresTables}.

\begin{table}[h!]
\begin{center}
\caption{The effect of treatment on subsequent number of users and edits}
\label{T:RegrUsersCrossSection}
\renewcommand{\tabcolsep}{3pt}
\begin{tabular}{ l cc cc cc cc }
\hline
  & \multicolumn{2}{c}{1st year} & \multicolumn{2}{c}{2nd year} & \multicolumn{2}{c}{3rd year} & \multicolumn{2}{c}{4th year} \\  & (1) & (2) & (3) & (4) & (5) & (6) & (7) & (8) \\ \hline & \multicolumn{8}{c}{Panel A. Dependent variable: average number of users per month} \\ 
Treatment group     &       0.099** &       0.122***&       0.096*  &       0.119***&      -0.006   &       0.014   &      -0.001   &       0.024   \\
                    &     (0.049)   &     (0.030)   &     (0.052)   &     (0.035)   &     (0.043)   &     (0.028)   &     (0.043)   &     (0.033)   \\
Language FE         &          No   &         Yes   &          No   &         Yes   &          No   &         Yes   &          No   &         Yes   \\
City FE             &          No   &         Yes   &          No   &         Yes   &          No   &         Yes   &          No   &         Yes   \\
Mean dep. var.      &       0.368   &       0.368   &       0.317   &       0.317   &       0.295   &       0.295   &       0.292   &       0.292   \\
SD dep. var.        &       0.329   &       0.329   &       0.353   &       0.353   &       0.287   &       0.287   &       0.285   &       0.285   \\
Adj. R-squared      &       0.017   &       0.678   &       0.013   &       0.608   &      -0.006   &       0.618   &      -0.006   &       0.453   \\
Observations        &         180   &         180   &         180   &         180   &         180   &         180   &         180   &         180   \\
 \hline & \multicolumn{8}{c}{Panel B. Dependent variable: average number of edits per month} \\ 
Treatment group     &       0.109** &       0.138***&       0.119*  &       0.140***&      -0.008   &       0.011   &       0.004   &       0.029   \\
                    &     (0.052)   &     (0.032)   &     (0.061)   &     (0.043)   &     (0.049)   &     (0.032)   &     (0.045)   &     (0.036)   \\
Language FE         &          No   &         Yes   &          No   &         Yes   &          No   &         Yes   &          No   &         Yes   \\
City FE             &          No   &         Yes   &          No   &         Yes   &          No   &         Yes   &          No   &         Yes   \\
Mean dep. var.      &       0.389   &       0.389   &       0.335   &       0.335   &       0.319   &       0.319   &       0.305   &       0.305   \\
SD dep. var.        &       0.351   &       0.351   &       0.412   &       0.412   &       0.325   &       0.325   &       0.302   &       0.302   \\
Adj. R-squared      &       0.019   &       0.663   &       0.015   &       0.556   &      -0.005   &       0.623   &      -0.006   &       0.424   \\
Observations        &         180   &         180   &         180   &         180   &         180   &         180   &         180   &         180   \\
 \hline & \multicolumn{8}{c}{Panel C. Dependent variable: average number of edits per month} \\ & \multicolumn{8}{c}{excluding edits of the text added by treatment} \\ 
Treatment group     &      -0.003   &       0.022   &       0.058   &       0.072*  &      -0.056   &      -0.035   &      -0.050   &      -0.026   \\
                    &     (0.048)   &     (0.030)   &     (0.059)   &     (0.043)   &     (0.048)   &     (0.031)   &     (0.043)   &     (0.033)   \\
Language FE         &          No   &         Yes   &          No   &         Yes   &          No   &         Yes   &          No   &         Yes   \\
City FE             &          No   &         Yes   &          No   &         Yes   &          No   &         Yes   &          No   &         Yes   \\
Mean dep. var.      &       0.333   &       0.333   &       0.305   &       0.305   &       0.295   &       0.295   &       0.278   &       0.278   \\
SD dep. var.        &       0.318   &       0.318   &       0.398   &       0.398   &       0.319   &       0.319   &       0.287   &       0.287   \\
Adj. R-squared      &      -0.006   &       0.641   &      -0.000   &       0.533   &       0.002   &       0.616   &       0.002   &       0.458   \\
Observations        &         180   &         180   &         180   &         180   &         180   &         180   &         180   &         180   \\

\hline
\end{tabular}
\end{center}
\footnotesize{
Notes: Each column presents estimates from a separate cross-section regression of 180 Wikipedia pages. Dependent variable is the average number of users or edits per month during the 1st year post-treatment (columns 1--2), 2nd (columns 3--4), 3rd (columns 5--6), and the 4th year post-treatment (columns 7--8). Dependent variables measure users (panel A), edits (panel B), and edits excluding those editing the text added by the treatment (panel C).
Standard errors are reported in parentheses. *** Indicates significance at 1 percent level, ** at 5 percent level, * at 10 percent level.
}
\end{table}

Panels A and B of \cref{T:RegrUsersCrossSection} show that the treatment increased the number of users and the number of edits during the first two years after the experiment.
Specifically, the treatment increased the average monthly number of users editing the page by about 0.13 users (columns 2 and 4 in panel A) and increased the average monthly number of edits by 0.14 edits (columns 2 and 4 in panel B). These effects are 30--40\% of a standard deviation.

\begin{figure}[h!]
\begin{center}
    \begin{subfigure}[b]{0.32\textwidth}
      \includegraphics[width=\textwidth]{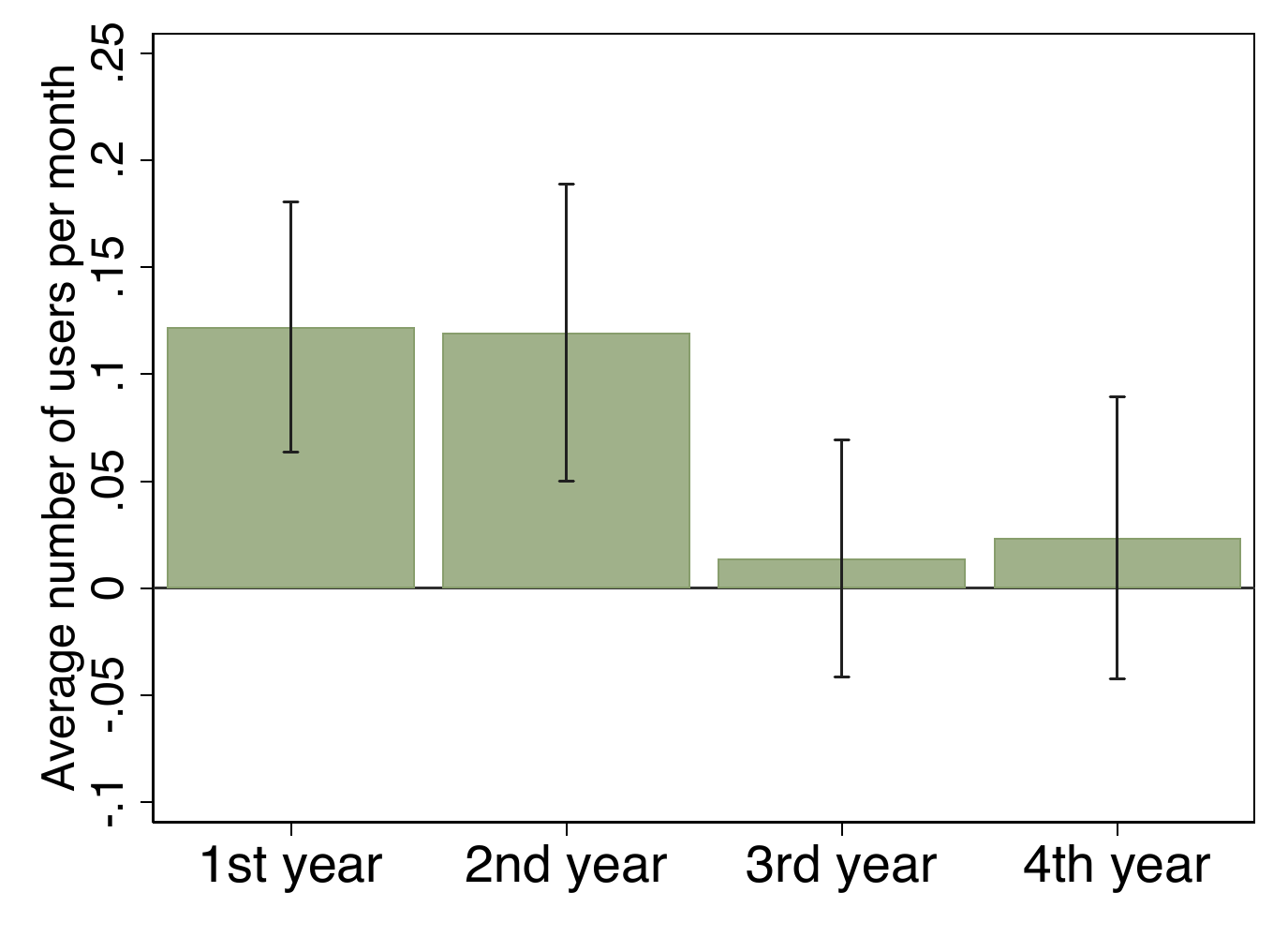}
      \caption{\# users}
      \label{F:RegrUsersCrossSection_users}
    \end{subfigure}
    \begin{subfigure}[b]{0.32\textwidth}
      \includegraphics[width=\textwidth]{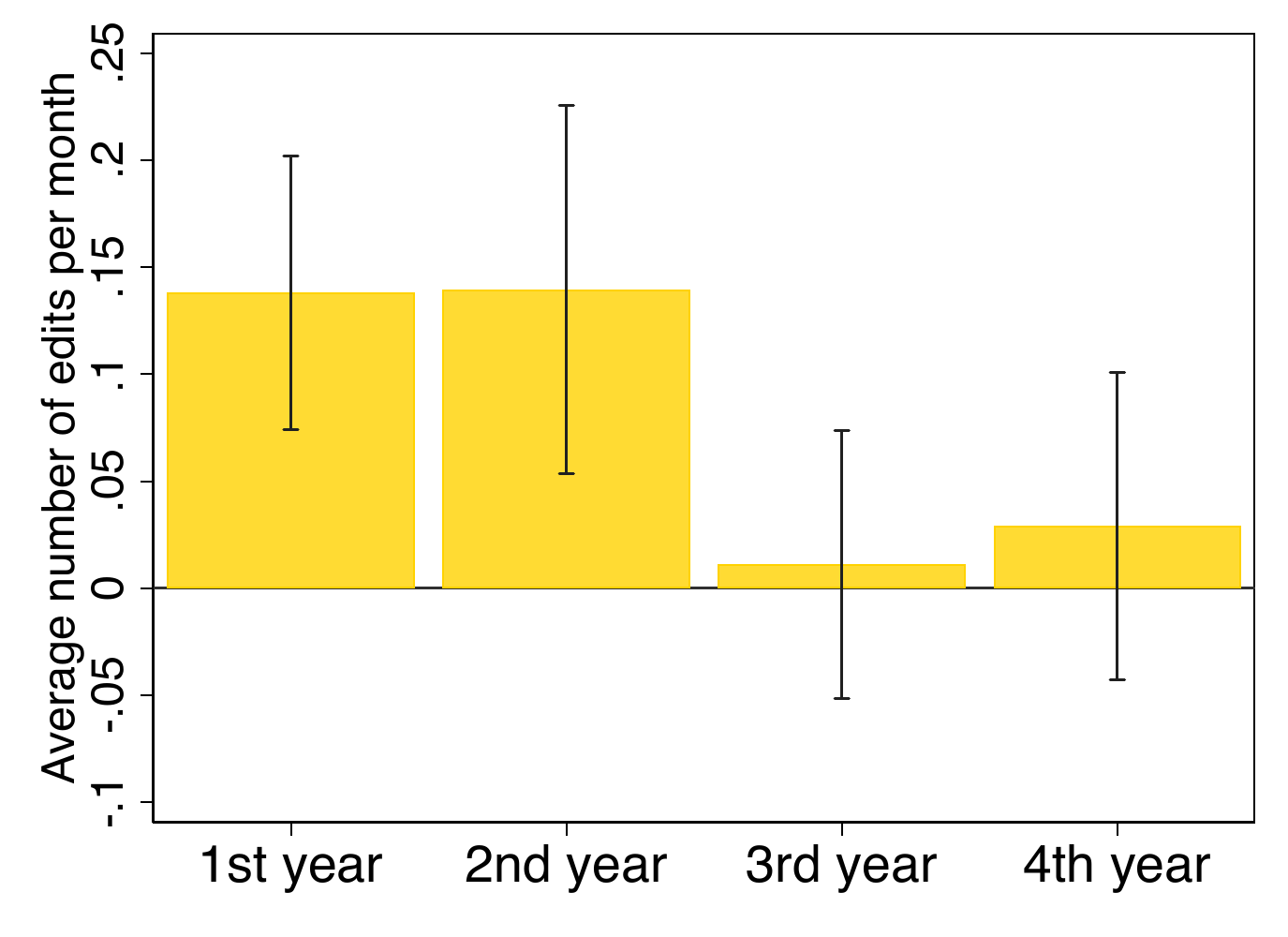}
      \caption{\# edits}
      \label{F:RegrUsersCrossSection_edits}
    \end{subfigure}
    \begin{subfigure}[b]{0.32\textwidth}
      \includegraphics[width=\textwidth]{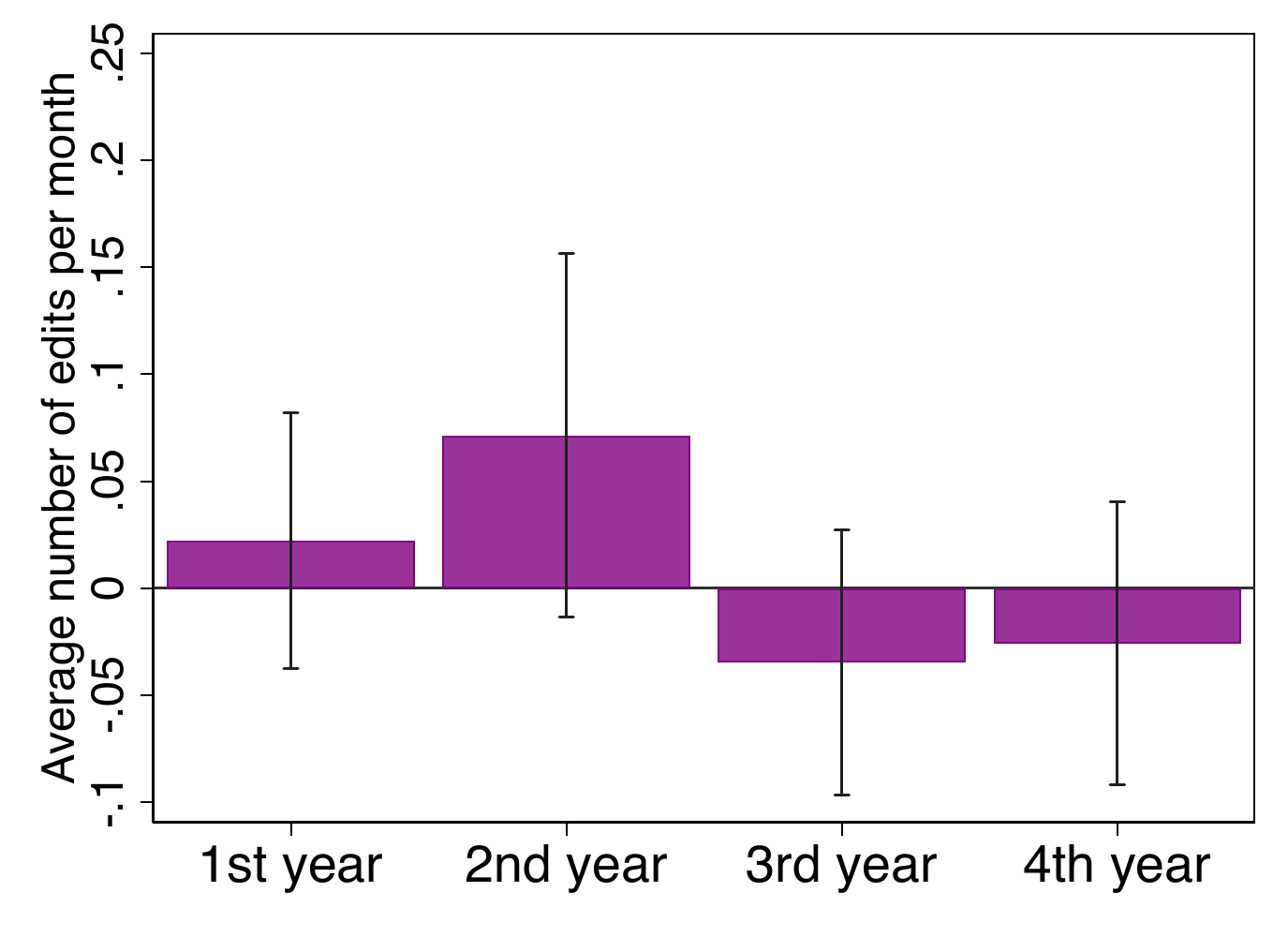}
      \caption{\small{\# edits excl. treatment}}
      \label{F:RegrUsersCrossSection_editsnotTT}
    \end{subfigure}
     \caption{The impact of treatment on the monthly average number of users and edits}
     \label{F:RegrUsersCrossSection}
\end{center}
     \footnotesize{
Notes: This figure presents point estimates (bars) and 95\% confidence intervals (lines) of coefficients from 12 regressions in columns 2, 4, 6, and 8 in panels A to C of \cref{T:RegrUsersCrossSection}. The coefficients describe the impact of treatment on the average monthly number of users (panel A), edits (panel B), and edits excluding those of the text added by the treatment (panel C) in the 1st, 2nd, 3rd, and 4th year post-treatment. Unit of observation is a page (180 pages). For further details see \cref{T:RegrUsersCrossSection}.
}
\end{figure}

However, these increases are only short-lived. 
In the third and fourth year, for both measures the effect of treatment is insignificant and the coefficients are small in magnitude. In the fourth year, the estimated increase is about 0.02 users per month, which is less than 9\% of a standard deviation, and in the third year, the effect is even smaller  (columns 6 and 8 in panels A and B). \Cref{F:RegrUsersCrossSection} summarizes these results. 

What do these editors and edits do in the first two years post-treatment if it has surprisingly little effect on page length (as we saw in \cref{S:ImpactOnOutcomes})? A natural explanation could be that the additional edits simply polish the text added by the treatment. To study this, we re-calculated the average number of edits per month while excluding edits that directly edited the text added by the treatment. The estimates using this outcome variable are presented in panel C of \cref{T:RegrUsersCrossSection} and \cref{F:RegrUsersCrossSection_editsnotTT}. The results show that when excluding the edits that directly affect the text added by the treatment then the treatment effect is much smaller. We conclude that a large share of the short-run increase in editing (seen in panels A and B of \cref{T:RegrUsersCrossSection}) comes from editing the content added by the treatment. 

\paragraph{Robustness and heterogeneity. }
In \cref{T:RegrRobustUsersGroupFE,T:RegrRobustUsersControlPastVar,T:RegrRobustUsersPanel} in online \cref{A:AdditionalFiguresTables} we assess whether our findings reported in panels A and B in \cref{T:RegrUsersCrossSection} hold under different specifications.
In brief, we find that the results are robust to including alternative covariates and estimating a panel data fixed effects model instead of cross-sectional regressions.

First, we assess whether the results are sensitive to the inclusion of alternative covariates. \Cref{T:RegrRobustUsersGroupFE} presents estimates from the same regressions as in \cref{T:RegrUsersCrossSection}, but instead of city and language fixed effects, it includes stratification group dummies (see \cref{S:Experiment}). Results are identical to those without any fixed effects in \cref{T:RegrUsersCrossSection}. Regressions in \cref{T:RegrRobustUsersControlPastVar} include pre-treatment average number of users and edits. The results remain similar to those in \cref{T:RegrUsersCrossSection}.

\Cref{T:RegrRobustUsersPanel} presents estimates from panel data fixed-effects regressions. The panel data fixed effects estimates are similar to those based on cross-sectional data. The treatment significantly increased the number of users and the number of edits in the first two years after the experiment but had no effect on the third and fourth year. The magnitude of the estimated effect in the first two years is only slightly larger compared to the cross-sectional framework: we estimate 0.14 users per month in the panel data regressions vs 0.12 in the cross-sectional regressions.

In \cref{T:RegrRobustUsers_fr,T:RegrRobustUsers_de,T:RegrRobustUsers_it} we re-estimate  regressions in \cref{T:RegrUsersCrossSection} separately by each language. We find that there is heterogeneity across languages. In particular, the treatment has a positive significant effect on editing behavior only in French and German Wikipedia and there is no evidence of a positive effect in Italian Wikipedia. 

To shed more light on where the short-run increase in editing activity comes from, we analyze the sample by splitting it in two ways. First, we distinguish edits by ``old'' users who have edited the same page prior to the experiment (i.e. users who might have a sense of ``ownership'' of the page) and ``new'' users who did not edit the page before the experiment. \Cref{T:RegrRobustEditsByOldUsersCS} shows no evidence that the treatment has any effect on old users. Instead, we find that the short-run increase in the number of users comes from new users. It should be noted that the number of old users is small, which might make it difficult to detect the effect. 

Second, we study the heterogeneous effects by the age of the Wikipedia pages. We hypothesized that in the early stages of page age, the effect of added content could be larger than in the later stages. We divided the sample by the median age of the page.\footnote{Note that the sample is split to unequal groups because many pages were created on a same day.} Estimates in \cref{T:RegrHeterUsersPageAge} suggest that, indeed, in the case of  younger pages the effect of treatment in the first year is larger. 

\subsection{The impact on edit distance}

\paragraph{Main results.}
\Cref{T:RegrEditDistCrossSection} presents the main results of the impact of treatment on the subsequent \emph{edit distance}, which equals the total amount of text added plus the total amount of text deleted. By adding together the amount of text added and deleted, edit distance captures how polished the text is and in this way provides a measure of quality.
As in the previous tables, each column presents estimates from a cross-sectional regression where the outcome variables measure editing activity in the 1st, 2nd, 3rd, and the 4th year after the experiment.
The dependent variable in panel A is the average edit distance per month. 

\begin{table}[h!]
\begin{center}
\caption{The effect of treatment on edit distance in the 4 years post-treatment}
\label{T:RegrEditDistCrossSection}
\renewcommand{\tabcolsep}{1pt}
\begin{tabular}{ l cc cc cc cc }
\hline
 & \multicolumn{2}{c}{1st year} & \multicolumn{2}{c}{2nd year} & \multicolumn{2}{c}{3rd year} & \multicolumn{2}{c}{4th year} \\  & (1) & (2) & (3) & (4) & (5) & (6) & (7) & (8) \\ \hline & \multicolumn{8}{c}{Panel A. Dependent variable: monthly average edit distance} \\ 
Treatment group     &       7.655   &      12.806   &     128.524   &     100.284   &     -15.897   &     -13.534   &      23.890   &       1.296   \\
                    &    (30.699)   &    (31.301)   &   (101.004)   &   (105.570)   &    (21.771)   &    (21.692)   &    (74.735)   &    (79.265)   \\
Language FE         &          No   &         Yes   &          No   &         Yes   &          No   &         Yes   &          No   &         Yes   \\
City FE             &          No   &         Yes   &          No   &         Yes   &          No   &         Yes   &          No   &         Yes   \\
Mean dep. var.      &     119.848   &     119.848   &     108.318   &     108.318   &      65.123   &      65.123   &     112.119   &     112.119   \\
SD dep. var.        &     205.394   &     205.394   &     678.724   &     678.724   &     145.853   &     145.853   &     500.082   &     500.082   \\
Adj. R-squared      &      -0.005   &       0.071   &       0.003   &       0.032   &      -0.003   &       0.115   &      -0.005   &      -0.005   \\
Observations        &         180   &         180   &         180   &         180   &         180   &         180   &         180   &         180   \\
  \hline & \multicolumn{8}{c}{Panel B. Dependent variable: monthly average capped edit distance} \\ 
Treatment group     &      13.644** &      17.282***&      18.642*  &      20.512** &      -6.174   &      -4.158   &      -2.823   &       0.485   \\
                    &     (6.124)   &     (5.609)   &     (9.776)   &     (8.654)   &     (6.812)   &     (5.294)   &     (6.140)   &     (5.343)   \\
Language FE         &          No   &         Yes   &          No   &         Yes   &          No   &         Yes   &          No   &         Yes   \\
City FE             &          No   &         Yes   &          No   &         Yes   &          No   &         Yes   &          No   &         Yes   \\
Mean dep. var.      &      39.173   &      39.173   &      35.008   &      35.008   &      32.183   &      32.183   &      28.758   &      28.758   \\
SD dep. var.        &      41.530   &      41.530   &      66.061   &      66.061   &      45.675   &      45.675   &      41.095   &      41.095   \\
Adj. R-squared      &       0.022   &       0.270   &       0.015   &       0.314   &      -0.001   &       0.463   &      -0.004   &       0.324   \\
Observations        &         180   &         180   &         180   &         180   &         180   &         180   &         180   &         180   \\

\hline
\end{tabular}
\end{center}
\footnotesize{
Notes: Each column presents estimates from a separate cross-section regression of 180 Wikipedia pages. Edit distance (panel A) equals the number of characters added plus the number of characters deleted. Capped edit distance (panel B) is calculated using individual edits which edit distance is capped at the 90th percentile. All dependent variables are averages across months and are measured in the number of characters.  Standard errors are reported in parentheses. *** Indicates significance at 1 percent level, ** at 5 percent level, * at 10 percent level.
}
\end{table}

Estimates in \cref{T:RegrEditDistCrossSection} panel A show that the treatment had no statistically significant effect on edit distance. Coefficients vary in sign and, with the exception of the second year,  are small in magnitude, less than 15 characters per month (columns 2, 6, and 8). In the second year, the (statistically insignificant) coefficient estimate (in column 4) implies the treatment effect of 100 characters per month, which is still slightly less than 15\% of a standard deviation.  

Because the distribution of edit distance has a long tail, in panel B of \cref{T:RegrEditDistCrossSection}, we use an alternative edit distance measure calculated from individual capped edits. The individual edits are capped from above at the 90th percentile. The 90th percentile equals about 500 characters and is about 10 times larger than the median edit. In this way, the capped edit distance measure gives a smaller weight to long edits. 

Estimates in panel B show that in the first two years after the experiment, the treatment increases the average capped edit distance. The magnitude of the effect is small---an increase of about 17 characters per month in the first year (column 2) and 21 characters per month in the second year (column 4). Provided that the average word length across languages in the experiment is about 10.4 characters, our treatment increased the edit distance by 1.5 words in the first and by two words in the second year.\footnote{Source: \url{http://www.ravi.io/language-word-lengths}.} The small magnitude of the short-term increase in edit distance is in line with the findings from panel C of \cref{T:RegrUsersCrossSection}, which showed that most of the increase in editing comes from editing the content added by the treatment.

The treatment has no significant effect on capped edit distance in later years. In the later years, the coefficients are even smaller. In the fourth year, the estimated effect is about 0.5 characters per month (column 8), which is less than 1\% of the standard deviation.

\paragraph{Robustness.}
\Cref{T:RegrEditDistCrossSection} already illustrated that the estimates of the impact of treatment on edit distance are sensitive to the specific form of the outcome variable. In the following, we assess whether the results remain similar under different specifications. First, we re-estimate the cross-sectional regressions including alternative covariates. Second, instead of using the cross-sectional data, we use a panel and estimate a panel data fixed effects model. In brief, the results in all these robustness checks are close to those in \cref{T:RegrEditDistCrossSection}. 

First, we assess whether the estimates are sensitive to including alternative covariates.
\Cref{T:RegrRobustEditDistGroupFE} presents estimates from the same regressions as in \cref{T:RegrEditDistCrossSection}, but instead of city and language fixed effects, it includes stratification group dummies (see \cref{S:Experiment}). Results are identical to those without any fixed effects in \cref{T:RegrEditDistCrossSection} (columns 1, 3, 5, and 7 in \cref{T:RegrEditDistCrossSection}). In \Cref{T:RegrRobustEditDistControlPastVar} we re-estimate the regressions in \cref{T:RegrEditDistCrossSection} including pre-treatment average edit distance measures as covariates. Note that the estimated coefficients on the added covariates have unstable signs and are not always statistically significant. But the estimates of the impact of treatment on edit distance remain similar to those in \cref{T:RegrEditDistCrossSection}.
  
\Cref{T:RegrRobustEditDistPanel} presents estimates from panel data fixed-effects regressions. These results are similar to those based on cross-sectional data. According to the estimates, the treatment increased capped net edit distance by about 18 characters per month in the first year post-treatment and 23 characters per month in the second year. It had no statistically significant effect in the following years, and the coefficients are very small, suggesting an effect of fewer than 2 characters per month.

Taken together, the results in \cref{T:RegrEditDistCrossSection,T:RegrRobustEditDistGroupFE,T:RegrRobustEditDistControlPastVar,T:RegrRobustEditDistPanel}  show no significant effect of treatment on the raw edit distance measure and show a small positive impact on the capped edit distance only in the first two years after the treatment.

\section{Discussion} \label{S:Discussion}

\paragraph{Comparison with previous literature.} \cite{aaltonen-seiler-2016} and \cite{nagaraj2017information} studied the same question and reached two opposite conclusions. \cite{aaltonen-seiler-2016} used detailed observational data from Wikipedia and found that longer pages get more future edits. Their simulation suggests a large cumulative effect---due to the positive externality, Wikipedia has almost twice as much content than without. On the other hand, \cite{nagaraj2017information} used quasi-random variation from a natural experiment in a Wikipedia-style mapping service. He found that regions with better initial content received fewer contributions and had worse quality output in the long run (about ten percent higher error rate).

In contrast to these papers, we used variation from a randomized field experiment, which provides clean identification of the causal impact. Our results bridge the gap between their opposing conclusions. We find that the addition of content has a negligible impact on the subsequent long-run growth of content. However, we do find small and short-lived increases in page length and editing activity. To be precise, coefficients estimated by \cite{aaltonen-seiler-2016} implied that an additional $10,000$ characters of text lead to $0.204$ additional users per week, which corresponds to about $0.18$ additional users per month for $2,000$  characters added by the treatment. Our estimates for the first two years after the experiment are slightly smaller, about $0.12$ additional users per month, and we find no significant long-run effects. Moreover, we also find that the short-run increase in editing activity is largely concentrated on the text added by the treatment. Taken together, our findings imply that in a setting where the variation comes from a randomized experiment the long-run externalities are negligible. 

We would like to note that differences in the results could be explained not only by differences in the identification methods, but also by subtle differences in the research settings. In the following, we highlight three such differences.
First, Wikipedia is a collaborative process, where it is not clear what the end product should look like. Each contribution can potentially signal to other editors the importance of a particular topic. For example, it is not clear which topics related to the Roman Empire (studied by \cite{aaltonen-seiler-2016}) are important enough to be covered in detail in Wikipedia. In this way, contributions become ``votes'' of importance. This effect is missing in the case of maps (studied by \cite{nagaraj2017information}), where the final output should be more or less homogeneous in the level of detail. We studied city pages, which have a standardized structure and therefore the vote-of-importance effect should be smaller than in \cite{aaltonen-seiler-2016}. Future research could further explore whether user-generated contributions also serve as signals or votes on the importance of a topic.

Second, in our setting, similar to \cite{nagaraj2017information}, the added content comes from an outside source and is not generated by the community itself.\footnote{Although, in our case, the added content is translated from the other language editions of Wikipedia.} Instead, in the setting of \cite{aaltonen-seiler-2016}, the content is created by the community. Content coming from an outside source might have a different impact than when it was generated by the existing users themselves.  

Third, the key mechanism proposed by \cite{nagaraj2017information} to explain the negative externality is the ``ownership effect'', which plays a less prominent role in Wikipedia. \cite{nagaraj2017information} suggested that contributors who added particular bridges or streets on the user-generated map may feel more responsible to keep these objects updated over time. Therefore, the treatment of adding more seeding information may backfire by not allowing the ownership of objects to arise naturally. Both \cite{aaltonen-seiler-2016} and our paper focus on textual content in Wikipedia, where ownership is less clear and we would thus expect the negative effect of adding content to be less prominent. Indeed, we find no evidence that old users (with a possible sense of ownership) change their editing behavior in response to the treatment, indicating that ownership plays a smaller role in Wikipedia editing. It would be interesting to see future research on the content ownership effect in a randomized experiment.

\paragraph{Implications.}
Many user-generated content platforms use managerial interventions that aim at motivating users to contribute new content. Examples include seeding the platform with initial content, compensating users for their contributions, or running campaigns to help to get the process started. Whether such policies should be used depends on whether the added content inspires an upward spiral of more user-generated content or whether it discourages future contributions. This choice is a critical managerial decision, because firm wikis, archives or Q\&A forums all depend on sufficient provision of information. On the other hand, such interventions are costly and require committing resources that cannot be invested elsewhere. Hence, it is important to understand not only the direction but also the magnitude of any possible externality that added content might have on follow-on contributions.

Our findings have a clear policy and managerial implication. The additional content has a negligible cumulative effect on growth. Therefore information seeding and incentivizing contributions are a matter of direct cost-benefit analysis: they pay off if and only if the costs of creating the content are lower than the value of the new content to the users and eventually to the platform. The additional costs or benefits via externalities that discourage or inspire future contributions are negligible. 

\paragraph{Limitations.}
Our analysis benefits from the clean identification due to the randomization, but it still faces limitations. The first concern is the generalizability of our results. 
As the above discussion already points out, the results from Wikipedia might not directly generalize to other  user-generated content platforms. One relevant difference among the platforms is the magnitude of the contributor's personal benefit. In Wikipedia, the personal benefit from contributing is likely to be smaller than in open-maps or open-source software. For example, a user of open-maps could directly benefit from correcting a mistake on a map, while a mistake in Wikipedia is unlikely to have any personal consequences. 

In this paper, we measured knowledge production using page length, number of users, number of edits, and edit distance. The same method and data can be used to estimate the impact on other measures, for example, number of images, number of external or internal links, or subjective quality measures.

\section{Conclusions} \label{S:Conclusions}

In this paper, we show that the addition of content has a negligible impact on the subsequent long-run growth of content. 
We identify the causal effect using exogenous variation from a randomized field experiment in Wikipedia. We find that the treatment, which added content to randomly chosen Wikipedia pages, had a negligible impact on the subsequent long-run growth. 
We do find some evidence of short-run increases in editing activity, in particular, increases in the number of edits and new editors in the first two years after the treatment. However, the amount of content these users added was small and most of their edits modified the content added by the treatment.
Our results are robust to a large set of sensitivity checks, including alternative covariates and analysis in the cross-sectional and in the panel data fixed effects framework. 

Our findings have a clear policy and managerial implication---information seeding and motivating content creation is not enough to generate a meaningful increase in future content generation. However, these policies are also not counterproductive as the discouragement effect on future contributions is also negligible. Therefore, it is simply a matter of direct cost-benefit analysis whether such policies pay off.

\phantomsection
\addcontentsline{toc}{section}{References}
\bibliography{mk-wikigrowth,toomash-wikigrowth,P_wikigrowth}
\bibliographystyle{econometrica}

\appendix
\counterwithin{table}{section}
\counterwithin{figure}{section}

\pagebreak
\setcounter{page}{1}\newpage
\renewcommand{\thepage}{A\arabic{page}}

\section{Online Appendix: Additional figures and tables} \label{A:AdditionalFiguresTables}
\renewcommand{\tablename}{Online Appendix Table}
\renewcommand{\figurename}{Online Appendix Figure}

\begin{figure}[!ht]
  \begin{center}
  \begin{subfigure}[b]{0.49\textwidth}
	\includegraphics[width=\textwidth]{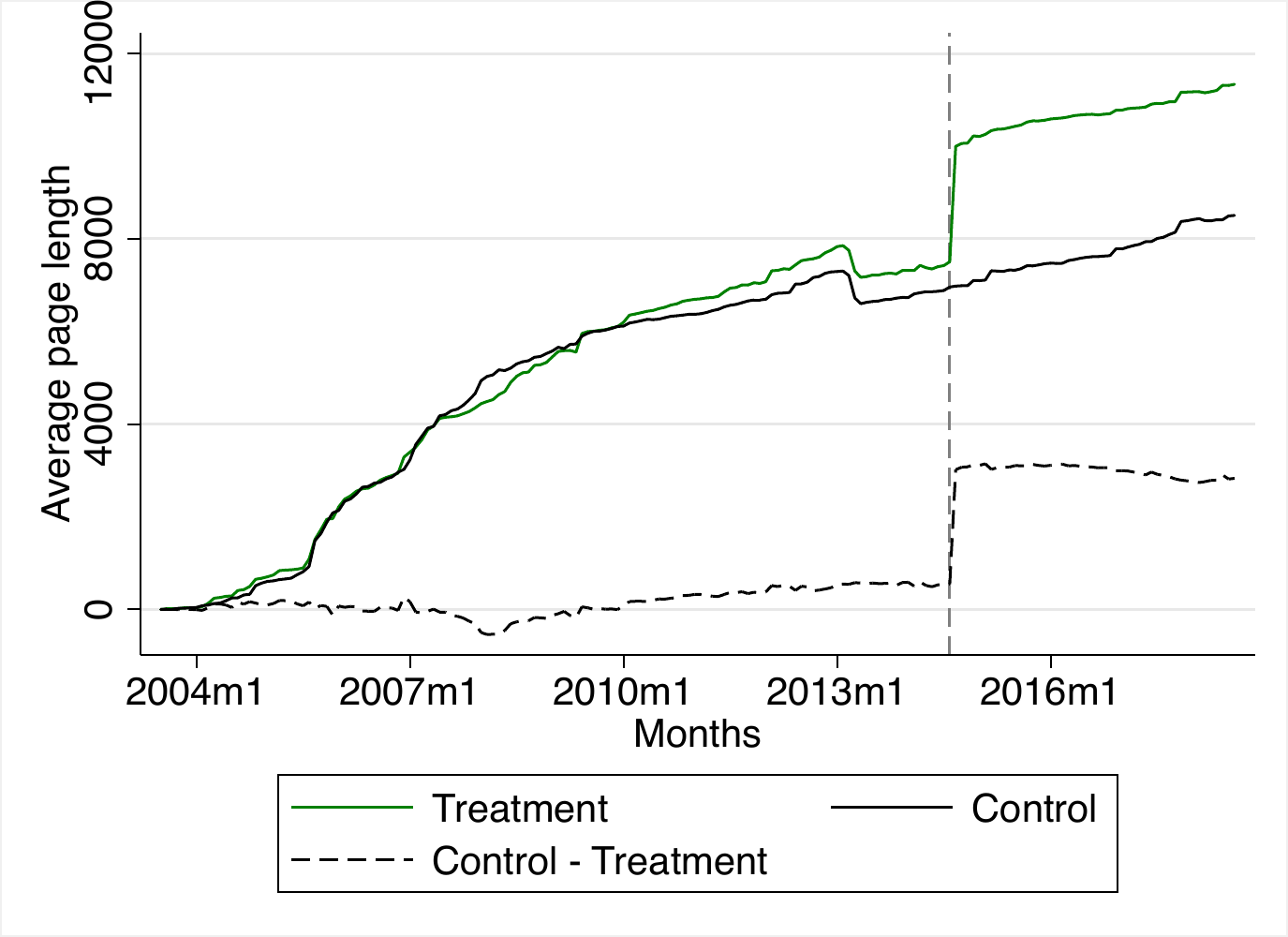}
  	\caption{Page length without Cordoba in French}
  	\label{F:Length_wo_FrenchCordoba}
  \end{subfigure} 
  \begin{subfigure}[b]{0.49\textwidth}
    	\includegraphics[width=\textwidth]{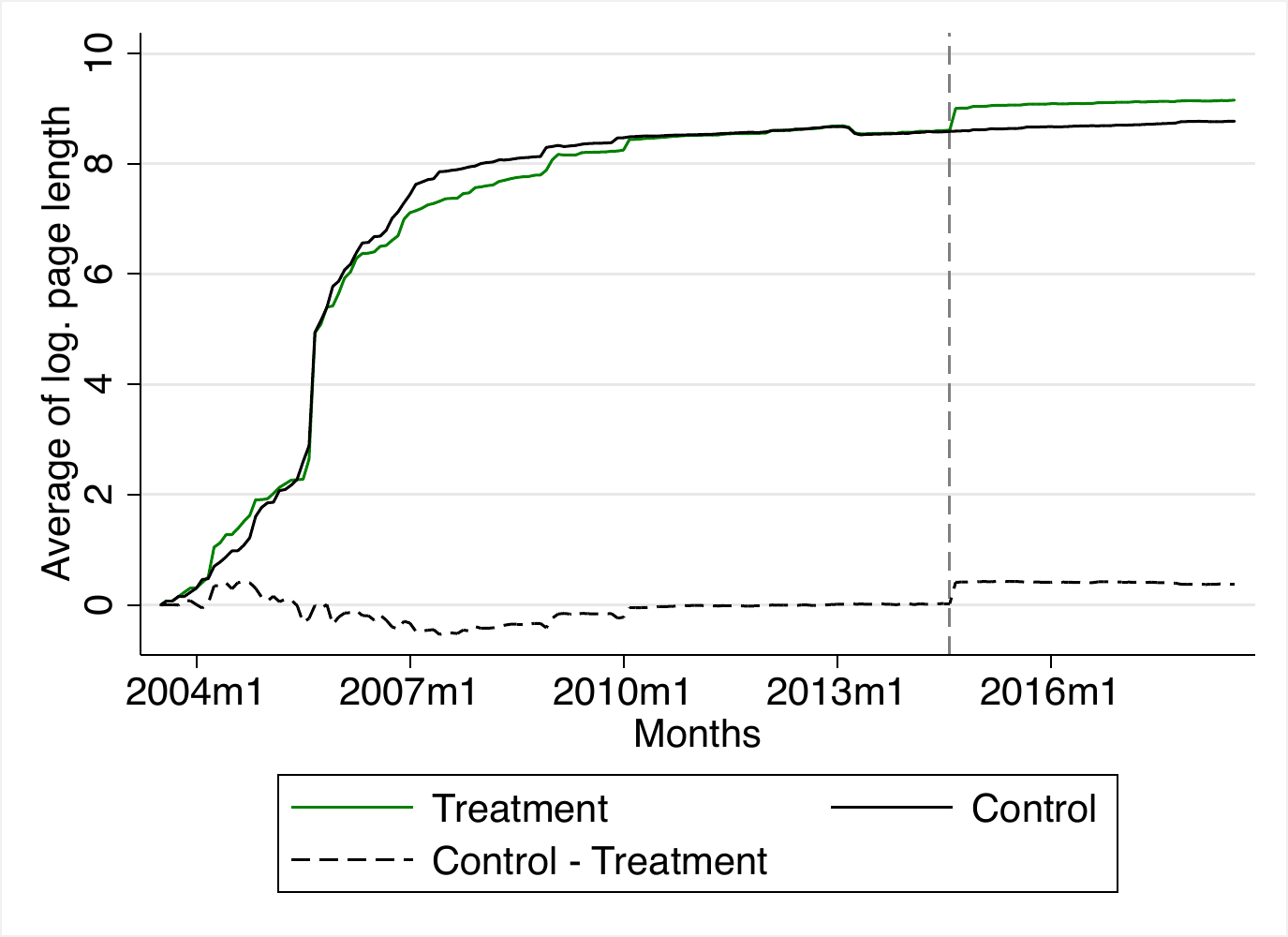}
      	\caption{Logarithm of page length}
    	\label{F:LogLength}
  \end{subfigure} 
  \caption{Robustness: Page length}
  \label{F:LengthOther}
  \end{center}
\footnotesize{
    Notes: The number of observations used to calculate the average is 90 in the control group and 89 (\Cref{F:Length_wo_FrenchCordoba}) or 90 (\Cref{F:LogLength}) in the treatment group. The experiment month (August 2014) is marked by dashed vertical line.
}   
\end{figure}

\begin{figure}[!ht]
  \begin{center}
  \begin{subfigure}[b]{0.49\textwidth}
	\includegraphics[width=\textwidth]{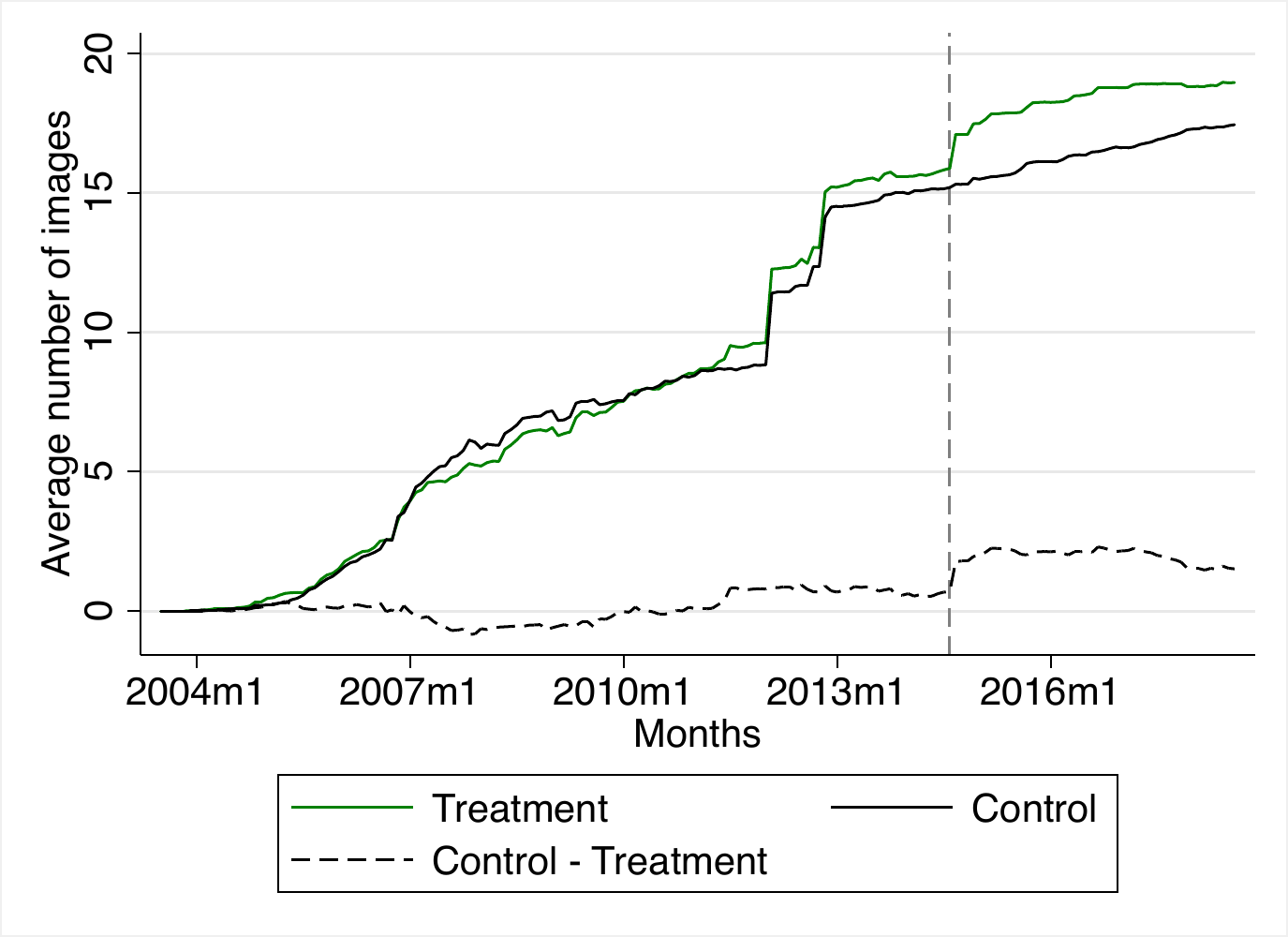}
  	\caption{Images}
  	\label{F:Images}
  \end{subfigure} 
  \begin{subfigure}[b]{0.49\textwidth}
    	\includegraphics[width=\textwidth]{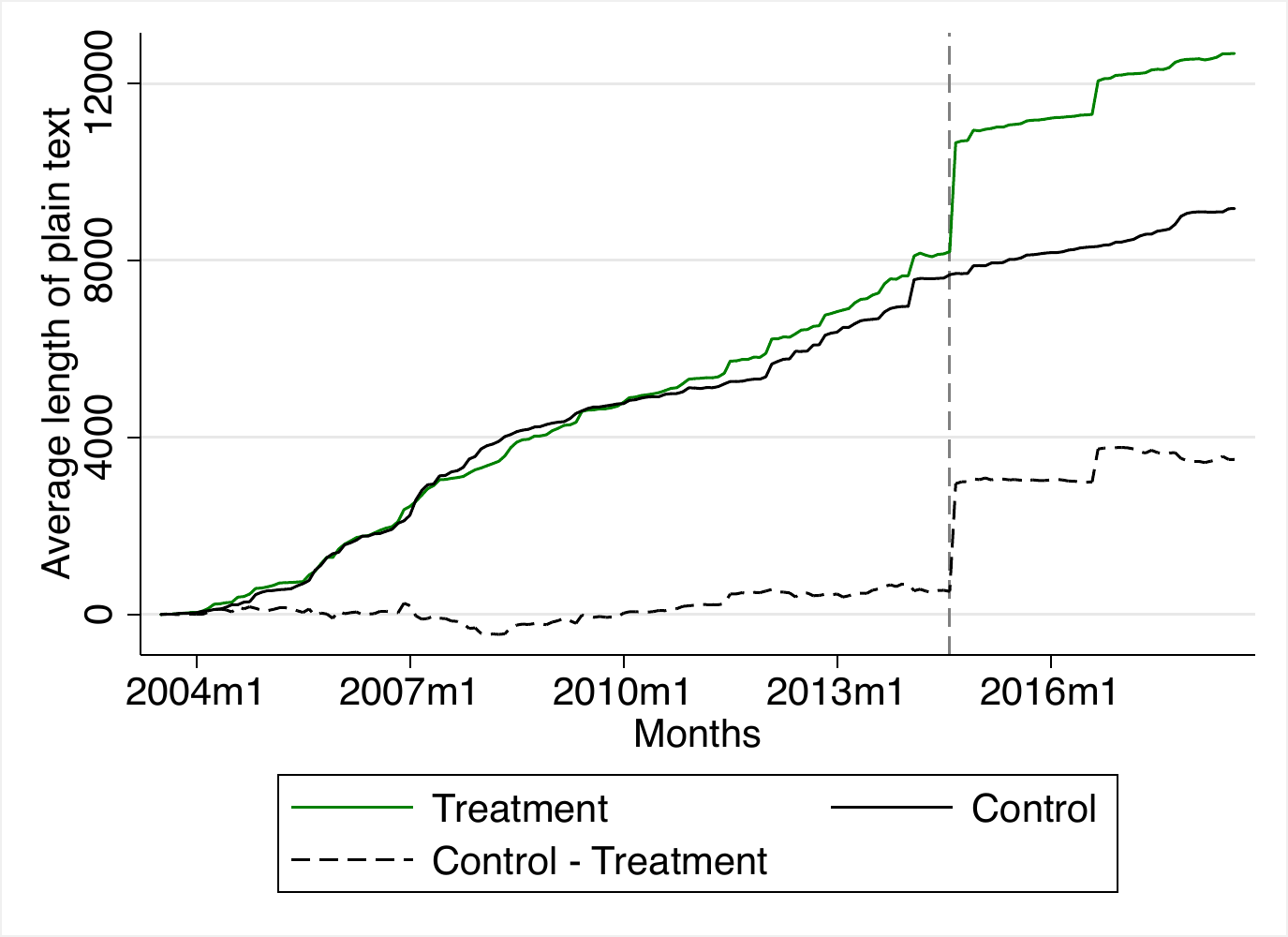}
      	\caption{Plain text}
    	\label{F:LengthPlainText}
  \end{subfigure} 
  \caption{Other output measures}
  \label{F:LengthImagesPlainText}
  \end{center}
\footnotesize{
    Notes: The number of observations is 90 in the control and 90 in the treatment groups. The experiment month (August 2014) is marked by dashed vertical line. \textit{Plain text} is obtained by removing html elements from the parsed text. 
} 
\end{figure}

\begin{figure}[!ht]
    \begin{center}
  \begin{subfigure}[b]{0.55\textwidth}
	\includegraphics[width=\textwidth]{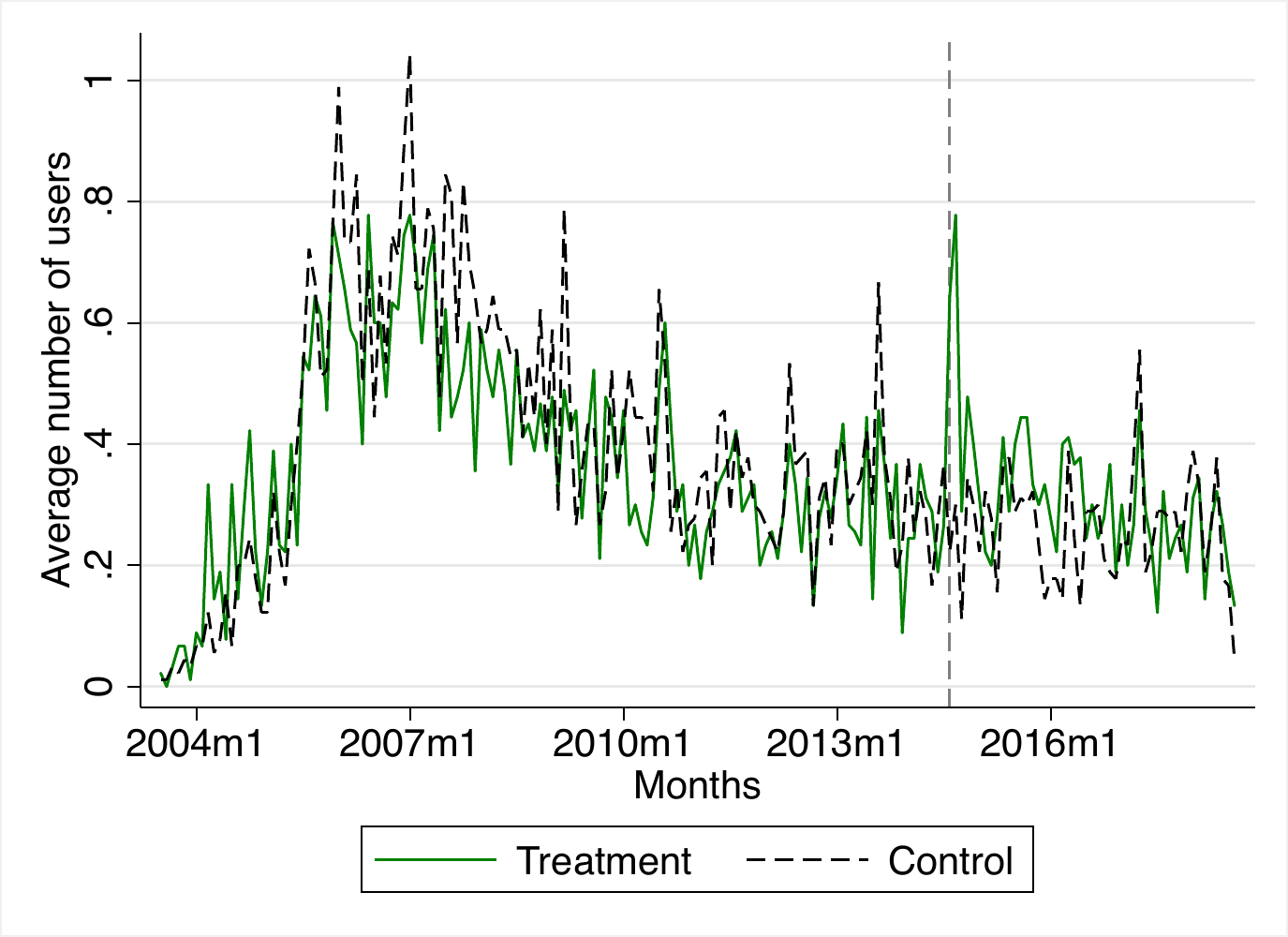}
  	\caption{Number of users}
  	\label{F:Users}
  \end{subfigure} 
  \begin{subfigure}[b]{0.55\textwidth}
	\includegraphics[width=\textwidth]{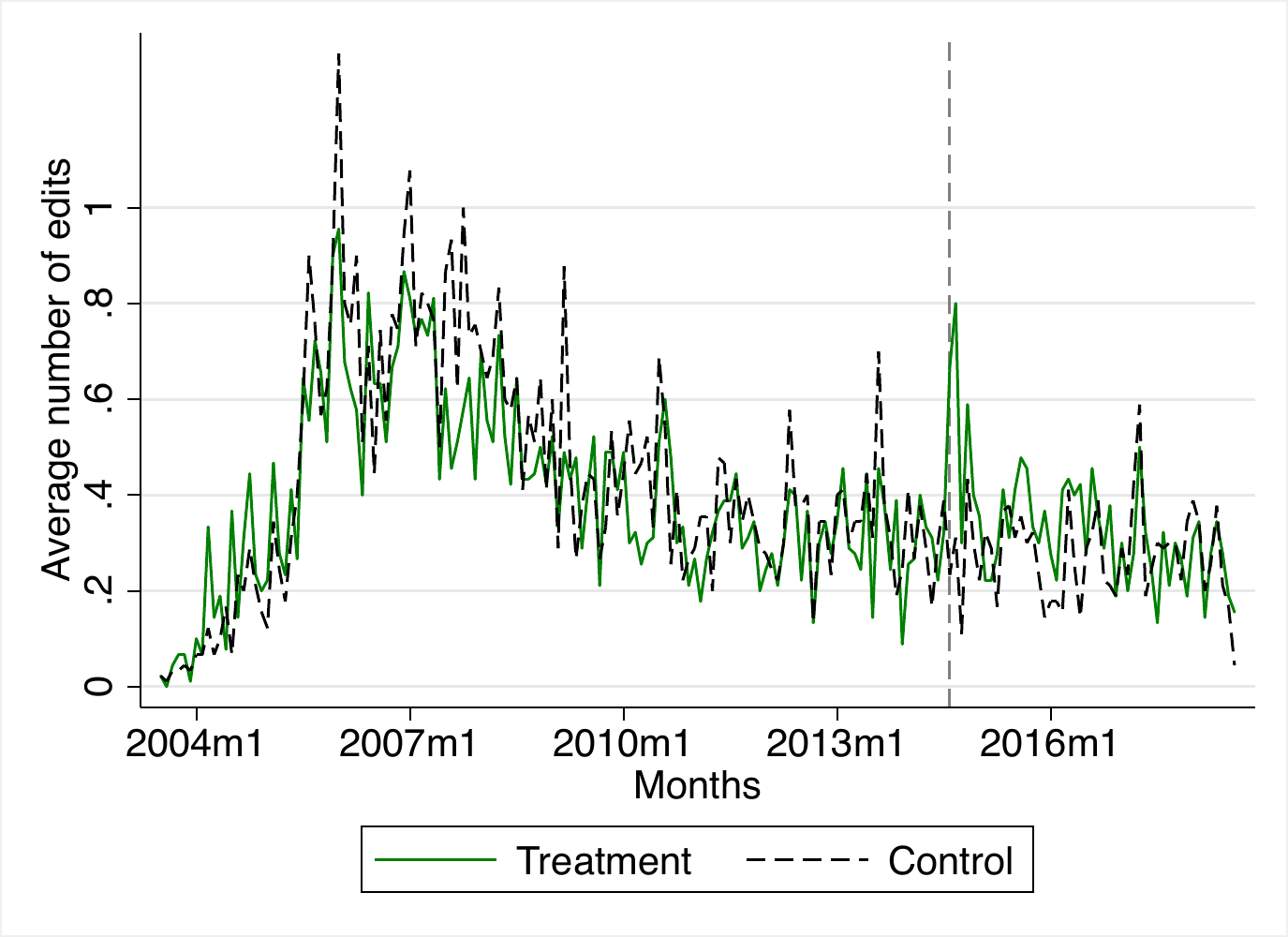}
  	\caption{Number of edits}
  	\label{F:EditDays}
  \end{subfigure} 
  \begin{subfigure}[b]{0.55\textwidth}
	\includegraphics[width=\textwidth]{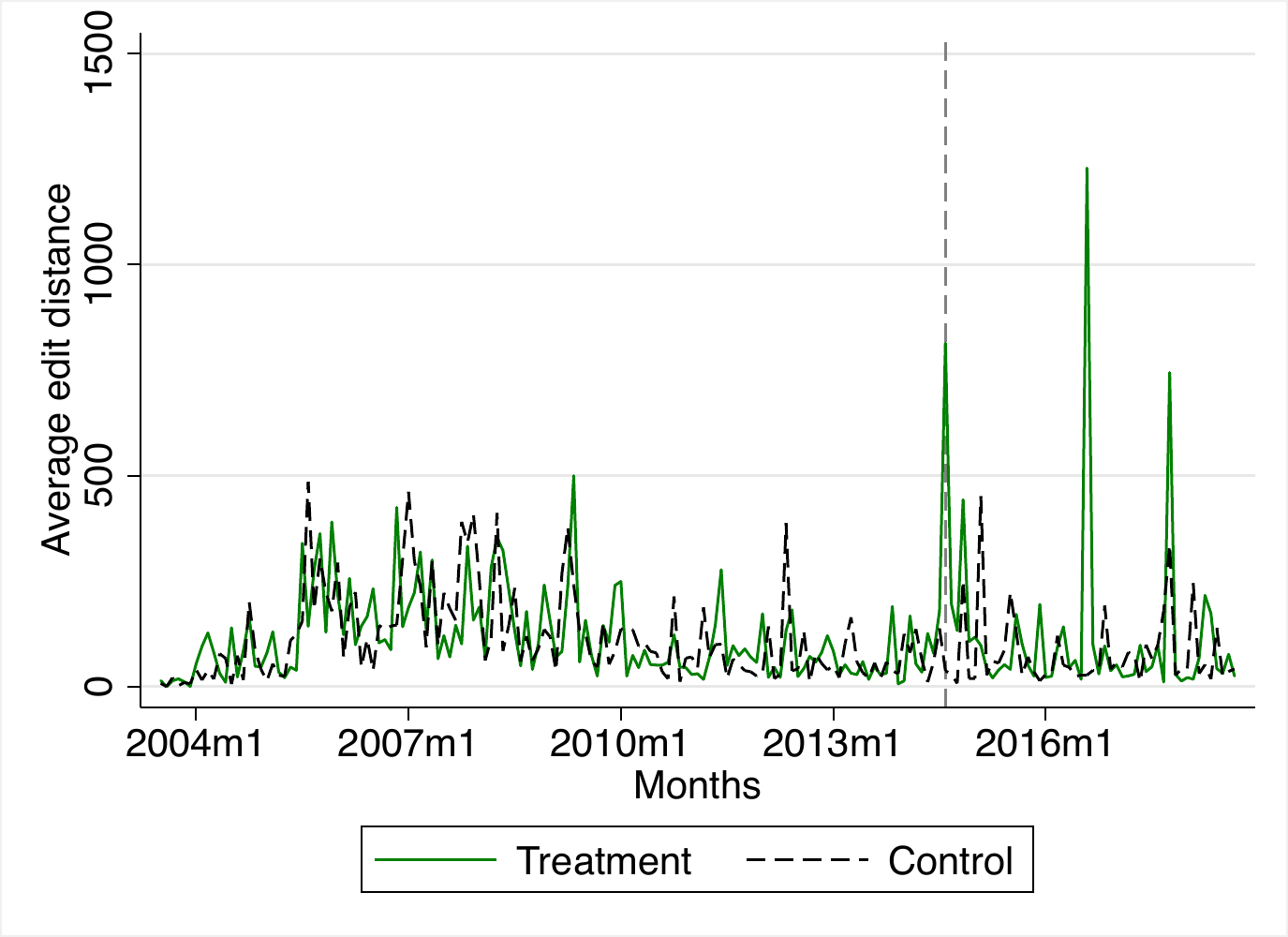}
  	\caption{Edit distance}
  	\label{F:EditDist}
  \end{subfigure} 
  \caption{Average input measures in the treatment and control groups per month}
  \label{F:EditingActivity}
  \end{center}
\footnotesize{
    Notes: The number of observations is 90 in the control and 90 in the treatment groups. The experiment month (August 2014) is marked by dashed vertical line.
}     
\end{figure}

\clearpage

\begin{table}[ht!]
\begin{center}
\caption{Comparison of pre-treatment characteristics in the treatment group versus the control group.}
\label{T:BalanceTest} 
\renewcommand{\tabcolsep}{3pt}
\begin{tabular}{ l ccccc} 
\hline
 & Control & Treatment & t-test & Wilcoxon test & Obs. \\ & group &  group & p-value & p-value & \\ & mean &  mean &  &  & \\  & (1) & (2) & (3) & (4) & (5) \\ \hline 
Log. length before treatment                                &    8.586&    8.611&    0.842&    0.655&      180\\
Aver. \# of users before treatment                          &    0.378&    0.333&    0.321&    0.285&      180\\
Aver. \# of edits before treatment                          &    0.396&    0.351&    0.341&    0.364&      180\\
Aver. edit dist. before treatment                           &   79.013&   76.786&    0.882&    0.738&      180\\
Aver. capped edit dist. before treatment                    &   42.511&   38.238&    0.470&    0.583&      180\\

\hline
\end{tabular}
\end{center}
\footnotesize{
Notes: 
Column 1 and 2 present the means of pre-treatment values of variables, separately for the control and the treatment group.  
Column 3 presents the p-value of the t-test for whether the difference between the control and treatment groups is significantly different from zero. Column 4 presents the p-value of the corresponding Wilcoxon rank-sum test. Column 5 presents the number of observations used in each test.
} 
\end{table}

\begin{table}[h!]
\begin{center}
\caption{Robustness: The effect of treatment on page length, alternative controls. Dependent variable: Logarithm of page length (minus treatment text).}
\label{T:RegrRobustLogLengthAlternControls} 
\renewcommand{\tabcolsep}{3pt}
\begin{tabular}{ l cc cc cc cc } 
\hline
  & \multicolumn{2}{c}{1st year} & \multicolumn{2}{c}{2nd year} & \multicolumn{2}{c}{3rd year} & \multicolumn{2}{c}{4th year} \\  & (1) & (2) & (3) & (4) & (5) & (6) & (7) & (8) \\ \hline  
Treatment group     &       0.068   &       0.045   &       0.071   &       0.047   &       0.059   &       0.035   &       0.040   &       0.017   \\
                    &     (0.115)   &     (0.036)   &     (0.122)   &     (0.044)   &     (0.123)   &     (0.048)   &     (0.126)   &     (0.056)   \\
Log. length before  &               &       0.936***&               &       0.968***&               &       0.969***&               &       0.959***\\
treatment           &               &     (0.024)   &               &     (0.029)   &               &     (0.032)   &               &     (0.037)   \\
Group FE            &         Yes   &         Yes   &         Yes   &         Yes   &         Yes   &         Yes   &         Yes   &         Yes   \\
Mean dep. var.      &       8.699   &       8.699   &       8.723   &       8.723   &       8.762   &       8.762   &       8.795   &       8.795   \\
SD dep. var.        &       0.792   &       0.792   &       0.836   &       0.836   &       0.847   &       0.847   &       0.868   &       0.868   \\
Adj. R-squared      &       0.046   &       0.908   &       0.048   &       0.876   &       0.048   &       0.855   &       0.057   &       0.811   \\
Observations        &         180   &         180   &         180   &         180   &         180   &         180   &         180   &         180   \\

\hline
\end{tabular}
\end{center}
\footnotesize{
Notes:
Each column presents estimates from a separate cross-section regression of 180 Wikipedia pages. The dependent variable is the logarithm of page length (minus the length of text added by the treatment) in the 1st year post-treatment (columns 1-2), 2nd (columns 3-4), 3rd (columns 5-6), and the 4th year post-treatment (columns 7-8). 
All regressions include stratification group dummies, \textit{GroupFE}. 
Standard errors are reported in parentheses. *** Indicates significance at 1 percent level, ** at 5 percent level, * at 10 percent level.
} 
\end{table}

\begin{table}[h!]
\begin{center}
\caption{The effect of treatment on page length, by language. Dependent variable: Logarithm of page length (minus treatment text).}
\label{T:RegrRobustLogLengthByLanguage} 
\renewcommand{\tabcolsep}{3pt}
\begin{tabular}{ l cc cc cc cc } 
\hline
  & \multicolumn{2}{c}{1st year} & \multicolumn{2}{c}{2nd year} & \multicolumn{2}{c}{3rd year} & \multicolumn{2}{c}{4th year} \\  & (1) & (2) & (3) & (4) & (5) & (6) & (7) & (8) \\ \hline & \multicolumn{8}{c}{Panel A. Sample: pages in French Wikipedia} \\ 
Treatment group     &       0.256   &       0.144*  &       0.328   &       0.210** &       0.285   &       0.168   &       0.223   &       0.109   \\
                    &     (0.201)   &     (0.080)   &     (0.216)   &     (0.096)   &     (0.219)   &     (0.105)   &     (0.224)   &     (0.125)   \\
Log. length before  &               &       0.807***&               &       0.850***&               &       0.842***&               &       0.821***\\
treatment           &               &     (0.046)   &               &     (0.055)   &               &     (0.060)   &               &     (0.071)   \\
Mean dep. var.      &       8.865   &       8.865   &       8.925   &       8.925   &       9.003   &       9.003   &       9.072   &       9.072   \\
SD dep. var.        &       0.782   &       0.782   &       0.847   &       0.847   &       0.852   &       0.852   &       0.869   &       0.869   \\
Adj. R-squared      &       0.011   &       0.843   &       0.021   &       0.808   &       0.012   &       0.772   &      -0.000   &       0.693   \\
Observations        &          60   &          60   &          60   &          60   &          60   &          60   &          60   &          60   \\
 \hline  & \multicolumn{8}{c}{Panel B. Sample: pages in German Wikipedia} \\ 
Treatment group     &       0.170   &       0.011   &       0.142   &      -0.015   &       0.145   &      -0.013   &       0.147   &      -0.007   \\
                    &     (0.160)   &     (0.047)   &     (0.160)   &     (0.050)   &     (0.161)   &     (0.051)   &     (0.157)   &     (0.053)   \\
Log. length before  &               &       0.942***&               &       0.936***&               &       0.939***&               &       0.913***\\
treatment           &               &     (0.038)   &               &     (0.040)   &               &     (0.041)   &               &     (0.043)   \\
Mean dep. var.      &       9.002   &       9.002   &       9.021   &       9.021   &       9.049   &       9.049   &       9.075   &       9.075   \\
SD dep. var.        &       0.621   &       0.621   &       0.618   &       0.618   &       0.621   &       0.621   &       0.609   &       0.609   \\
Adj. R-squared      &       0.002   &       0.914   &      -0.004   &       0.904   &      -0.003   &       0.900   &      -0.002   &       0.887   \\
Observations        &          60   &          60   &          60   &          60   &          60   &          60   &          60   &          60   \\
 \hline  & \multicolumn{8}{c}{Panel C. Sample: pages in Italian Wikipedia} \\ 
Treatment group     &      -0.223   &       0.008   &      -0.258   &      -0.016   &      -0.253   &      -0.011   &      -0.249   &       0.000   \\
                    &     (0.193)   &     (0.009)   &     (0.204)   &     (0.034)   &     (0.205)   &     (0.038)   &     (0.213)   &     (0.047)   \\
Log. length before  &               &       0.989***&               &       1.032***&               &       1.034***&               &       1.064***\\
treatment           &               &     (0.006)   &               &     (0.023)   &               &     (0.025)   &               &     (0.031)   \\
Mean dep. var.      &       8.231   &       8.231   &       8.222   &       8.222   &       8.233   &       8.233   &       8.238   &       8.238   \\
SD dep. var.        &       0.751   &       0.751   &       0.795   &       0.795   &       0.798   &       0.798   &       0.827   &       0.827   \\
Adj. R-squared      &       0.006   &       0.998   &       0.010   &       0.973   &       0.009   &       0.967   &       0.006   &       0.952   \\
Observations        &          60   &          60   &          60   &          60   &          60   &          60   &          60   &          60   \\

\hline
\end{tabular}
\end{center}
\footnotesize{
Notes:
Each column presents estimates from a separate cross-section regression of 60 Wikipedia pages. The dependent variable is the logarithm of page length (minus the length of text added by the treatment) in the 1st year post-treatment (columns 1-2), 2nd (columns 3-4), 3rd (columns 5-6), and the 4th year post-treatment (columns 7-8). 
Sample is restricted to pages in French Wikipedia (panel A), pages in German Wikipedia (panel B) or pages in Italian Wikipedia (panel C).
Standard errors are reported in parentheses. *** Indicates significance at 1 percent level, ** at 5 percent level, * at 10 percent level. 
} 
\end{table}

\begin{table}[h!]
\begin{center}
\caption{The effect of treatment on page length, by page age. Dependent variable: Logarithm of page length (minus treatment text).}
\label{T:RegrHeterLogLengthPageAge} 
\renewcommand{\tabcolsep}{2pt}
\begin{tabular}{ l cc cc cc cc } 
\hline
  & \multicolumn{2}{c}{1st year} & \multicolumn{2}{c}{2nd year} & \multicolumn{2}{c}{3rd year} & \multicolumn{2}{c}{4th year} \\  & (1) & (2) & (3) & (4) & (5) & (6) & (7) & (8) \\ \hline & \multicolumn{8}{c}{Panel A. Sample: pages with age below median} \\ 
Treatment group     &       0.239   &       0.109   &       0.244   &       0.114   &       0.237   &       0.114   &       0.142   &       0.026   \\
                    &     (0.188)   &     (0.085)   &     (0.189)   &     (0.084)   &     (0.186)   &     (0.091)   &     (0.190)   &     (0.110)   \\
Log. length before  &               &       0.825***&               &       0.831***&               &       0.796***&               &       0.760***\\
treatment           &               &     (0.054)   &               &     (0.054)   &               &     (0.058)   &               &     (0.070)   \\
Language FE         &          No   &         Yes   &          No   &         Yes   &          No   &         Yes   &          No   &         Yes   \\
Mean dep. var.      &       8.638   &       8.638   &       8.660   &       8.660   &       8.710   &       8.710   &       8.761   &       8.761   \\
SD dep. var.        &       0.732   &       0.732   &       0.735   &       0.735   &       0.724   &       0.724   &       0.734   &       0.734   \\
Adj. R-squared      &       0.010   &       0.800   &       0.011   &       0.807   &       0.010   &       0.765   &      -0.008   &       0.666   \\
Observations        &          60   &          60   &          60   &          60   &          60   &          60   &          60   &          60   \\
 \hline  & \multicolumn{8}{c}{Panel B. Sample: pages with age above median} \\ 
Treatment group     &      -0.018   &       0.029   &      -0.016   &       0.030   &      -0.030   &       0.015   &      -0.010   &       0.034   \\
                    &     (0.151)   &     (0.020)   &     (0.162)   &     (0.039)   &     (0.166)   &     (0.041)   &     (0.170)   &     (0.045)   \\
Log. length before  &               &       0.978***&               &       1.009***&               &       1.010***&               &       1.020***\\
treatment           &               &     (0.016)   &               &     (0.030)   &               &     (0.031)   &               &     (0.035)   \\
Language FE         &          No   &         Yes   &          No   &         Yes   &          No   &         Yes   &          No   &         Yes   \\
Mean dep. var.      &       8.730   &       8.730   &       8.754   &       8.754   &       8.788   &       8.788   &       8.812   &       8.812   \\
SD dep. var.        &       0.822   &       0.822   &       0.883   &       0.883   &       0.904   &       0.904   &       0.930   &       0.930   \\
Adj. R-squared      &      -0.008   &       0.982   &      -0.008   &       0.941   &      -0.008   &       0.940   &      -0.008   &       0.931   \\
Observations        &         120   &         120   &         120   &         120   &         120   &         120   &         120   &         120   \\

\hline
\end{tabular}
\end{center}
\footnotesize{
Notes:
Each column presents estimates from a separate cross-section regression of  Wikipedia pages. The dependent variable is the logarithm of page length (minus the length of text added by the treatment) in the 1st year post-treatment (columns 1-2), 2nd (columns 3-4), 3rd (columns 5-6), and the 4th year post-treatment (columns 7-8). 
Sample is restricted to pages with age below median (panel A) or above median (panel B). The sample is split to unequal groups because many pages were created on a same day. 
Standard errors are reported in parentheses. *** Indicates significance at 1 percent level, ** at 5 percent level, * at 10 percent level.
} 
\end{table}

\clearpage

\begin{table}[h!]
\begin{center}
\caption{Robustness: The effect of treatment on subsequent number of users and edits, alternative controls}
\label{T:RegrRobustUsersGroupFE} 
\renewcommand{\tabcolsep}{9pt}
\begin{tabular}{ l cccc } 
\hline
  & 1st year & 2nd year & 3rd year & 4th year \\  & (1) & (2) & (3) & (4) \\ \hline  \multicolumn{5}{c}{Panel A. Dependent variable: average number of users per month} \\ 
Treatment group     &       0.099** &       0.096*  &      -0.006   &      -0.001   \\
                    &     (0.047)   &     (0.051)   &     (0.042)   &     (0.041)   \\
Group FE            &         Yes   &         Yes   &         Yes   &         Yes   \\
Mean dep. var.      &       0.368   &       0.317   &       0.295   &       0.292   \\
SD dep. var.        &       0.329   &       0.353   &       0.287   &       0.285   \\
Adj. R-squared      &       0.067   &       0.058   &       0.024   &       0.070   \\
Observations        &         180   &         180   &         180   &         180   \\
 \hline  \multicolumn{5}{c}{Panel B. Dependent variable: average number of edits per month} \\ 
Treatment group     &       0.109** &       0.119*  &      -0.008   &       0.004   \\
                    &     (0.051)   &     (0.060)   &     (0.048)   &     (0.043)   \\
Group FE            &         Yes   &         Yes   &         Yes   &         Yes   \\
Mean dep. var.      &       0.389   &       0.335   &       0.319   &       0.305   \\
SD dep. var.        &       0.351   &       0.412   &       0.325   &       0.302   \\
Adj. R-squared      &       0.067   &       0.041   &       0.022   &       0.072   \\
Observations        &         180   &         180   &         180   &         180   \\

\hline
\end{tabular}
\end{center}
\footnotesize{
Notes: 
Notes: Each column presents estimates from a separate cross-section regression of 180 Wikipedia pages. Dependent variable is the average number of users (panel A) or edits (panel B) per month during the 1st year post-treatment (column 1), 2nd (column 2), 3rd (column 3), and the 4th year post-treatment (column 4). 
All regressions include stratification group dummies, \textit{Group FE}. 
Standard errors are reported in parentheses. *** Indicates significance at 1 percent level, ** at 5 percent level, * at 10 percent level. 
} 
\end{table}

\begin{table}[h!]
\begin{center}
\caption{Robustness: The effect of treatment on subsequent number of users and edits, past measures as controls}
\label{T:RegrRobustUsersControlPastVar} 
\renewcommand{\tabcolsep}{2pt}
\begin{tabular}{ l cccc cccc} 
\hline
  & \multicolumn{2}{c}{1st year} & \multicolumn{2}{c}{2nd year} & \multicolumn{2}{c}{3rd year} & \multicolumn{2}{c}{4th year} \\  & (1) & (2) & (3) & (4) & (5) & (6) & (7) & (8) \\ \hline & \multicolumn{8}{c}{Panel A. Dependent variable: average number of users per month} \\ 
Treatment group     &       0.136***&       0.136***&       0.134***&       0.134***&       0.026   &       0.021   &       0.029   &       0.037   \\
                    &     (0.032)   &     (0.027)   &     (0.036)   &     (0.033)   &     (0.030)   &     (0.028)   &     (0.030)   &     (0.031)   \\
Aver. \# of users   &       0.813***&       0.477***&       0.841***&       0.509***&       0.693***&       0.233** &       0.667***&       0.451***\\
before treatment    &     (0.053)   &     (0.095)   &     (0.060)   &     (0.115)   &     (0.049)   &     (0.097)   &     (0.050)   &     (0.111)   \\
Language FE         &          No   &         Yes   &          No   &         Yes   &          No   &         Yes   &          No   &         Yes   \\
City FE             &          No   &         Yes   &          No   &         Yes   &          No   &         Yes   &          No   &         Yes   \\
Mean dep. var.      &       0.368   &       0.368   &       0.317   &       0.317   &       0.295   &       0.295   &       0.292   &       0.292   \\
SD dep. var.        &       0.329   &       0.329   &       0.353   &       0.353   &       0.287   &       0.287   &       0.285   &       0.285   \\
Adj. R-squared      &       0.573   &       0.733   &       0.529   &       0.661   &       0.526   &       0.632   &       0.493   &       0.517   \\
Observations        &         180   &         180   &         180   &         180   &         180   &         180   &         180   &         180   \\
 \hline & \multicolumn{8}{c}{Panel B. Dependent variable: average number of edits per month} \\ 
Treatment group     &       0.145***&       0.150***&       0.158***&       0.154***&       0.025   &       0.018   &       0.033   &       0.041   \\
                    &     (0.036)   &     (0.030)   &     (0.044)   &     (0.041)   &     (0.034)   &     (0.031)   &     (0.034)   &     (0.035)   \\
Aver. \# of edits   &       0.795***&       0.417***&       0.887***&       0.514***&       0.748***&       0.237** &       0.646***&       0.420***\\
before treatment    &     (0.057)   &     (0.098)   &     (0.070)   &     (0.133)   &     (0.053)   &     (0.100)   &     (0.053)   &     (0.112)   \\
Language FE         &          No   &         Yes   &          No   &         Yes   &          No   &         Yes   &          No   &         Yes   \\
City FE             &          No   &         Yes   &          No   &         Yes   &          No   &         Yes   &          No   &         Yes   \\
Mean dep. var.      &       0.389   &       0.389   &       0.335   &       0.335   &       0.319   &       0.319   &       0.305   &       0.305   \\
SD dep. var.        &       0.351   &       0.351   &       0.412   &       0.412   &       0.325   &       0.325   &       0.302   &       0.302   \\
Adj. R-squared      &       0.530   &       0.706   &       0.478   &       0.603   &       0.522   &       0.638   &       0.449   &       0.482   \\
Observations        &         180   &         180   &         180   &         180   &         180   &         180   &         180   &         180   \\

\hline
\end{tabular}
\end{center}
\footnotesize{
Notes: Each column presents estimates from a separate cross-section regression of 180 Wikipedia pages. Dependent variable is the average number of users (panel A) or edits (panel B) per month during the 1st year post-treatment (columns 1-2), 2nd (columns 3-4), 3rd (columns 5-6), and the 4th year post-treatment (columns 7-8). 
All regressions include as a covariate the average number of users or edits pre-treatment.
Standard errors are reported in parentheses *** Indicates significance at the 1 percent level, ** at 5 percent level, * at 10 percent level.
} 
\end{table}

\begin{table}[h!]
\begin{center}
\caption{Robustness: The effect of treatment on subsequent number of users and edits, panel data FE regression}
\label{T:RegrRobustUsersPanel} 
\renewcommand{\tabcolsep}{4pt}
\begin{tabular}{ l cc cc } 
\hline
 & \multicolumn{2}{c}{\# users} & \multicolumn{2}{c}{\# edits}  \\ & (1) & (2) & (3) & (4) \\ \hline 
Treatment group,    &       0.144***&       0.144***&       0.154***&       0.154***\\
post-treatment 1st year&     (0.033)   &     (0.033)   &     (0.037)   &     (0.037)   \\
Treatment group,    &       0.141***&       0.141***&       0.164***&       0.164***\\
post-treatment 2nd year&     (0.037)   &     (0.036)   &     (0.045)   &     (0.044)   \\
Treatment group,    &       0.039   &       0.039   &       0.037   &       0.037   \\
post-treatment 3rd year&     (0.032)   &     (0.032)   &     (0.036)   &     (0.035)   \\
Treatment group,    &       0.044   &       0.044   &       0.049   &       0.049   \\
post-treatment 4th year&     (0.034)   &     (0.033)   &     (0.037)   &     (0.037)   \\
Month FE            &         Yes   &         Yes   &         Yes   &         Yes   \\
Month-language FE   &          No   &         Yes   &          No   &         Yes   \\
Page FE             &         Yes   &         Yes   &         Yes   &         Yes   \\
Mean dep. var.      &       0.338   &       0.338   &       0.356   &       0.356   \\
SD dep. var.        &       0.706   &       0.706   &       0.773   &       0.773   \\
Observations        &       18360   &       18360   &       18360   &       18360   \\

\hline
\end{tabular}
\end{center}
\footnotesize{
Notes: A unit of observation is a page-month pair. 
Dependent variable is the number of users (columns 1-2) or edits (columns 3-4) per month. 
All regressions include page fixed effects and either month fixed effects or month-language fixed effects. 
\textit{Treatment group, post-treatment 1st year} is an indicator variable that takes value one during the first year post-treatment if the page belongs to the treatment group and zero otherwise;  and similarly for other years. 
The sample is a balanced sample that starts from the time period when all the 180 pages existed. 
Standard errors, reported in parentheses, are clustered by page (180 pages). *** Indicates significance at the 1 percent level, ** at 5 percent level, * at 10 percent level.
} 
\end{table}

\begin{table}[htb]
\begin{center}
\caption{The effect of treatment on subsequent number of users  and edits, pages in French Wikipedia}
\label{T:RegrRobustUsers_fr}
\renewcommand{\tabcolsep}{9pt}
\begin{tabular}{ l cc cc }
\hline
  & 1st year & 2nd year & 3rd year & 4th year \\  & (1) & (2) & (3) & (4) \\ \hline \multicolumn{5}{c}{Panel A. Dependent variable: average number of users per month} \\ 
Treatment group     &       0.139   &       0.183   &      -0.017   &      -0.003   \\
                    &     (0.091)   &     (0.113)   &     (0.082)   &     (0.080)   \\
Mean dep. var.      &       0.392   &       0.372   &       0.325   &       0.326   \\
SD dep. var.        &       0.357   &       0.443   &       0.315   &       0.306   \\
Adj. R-squared      &       0.022   &       0.027   &      -0.017   &      -0.017   \\
Observations        &          60   &          60   &          60   &          60   \\
 \hline  \multicolumn{5}{c}{Panel B. Dependent variable: average number of edits per month} \\ 
Treatment group     &       0.147   &       0.244*  &      -0.028   &      -0.003   \\
                    &     (0.093)   &     (0.141)   &     (0.096)   &     (0.081)   \\
Mean dep. var.      &       0.429   &       0.408   &       0.353   &       0.332   \\
SD dep. var.        &       0.367   &       0.557   &       0.371   &       0.312   \\
Adj. R-squared      &       0.024   &       0.033   &      -0.016   &      -0.017   \\
Observations        &          60   &          60   &          60   &          60   \\

\hline
\end{tabular}
\end{center}
\footnotesize{
Notes: Each column presents estimates from a separate cross-section regression of 60 Wikipedia pages in French Wikipedia. 
Dependent variable is the average number of users (panel A) or edits (panel B) per month during the 1st year post-treatment (column 1), 2nd (column 2), 3rd (column 3), and the 4th year post-treatment (column 4). 
Standard errors are reported in parentheses. *** Indicates significance at the 1 percent level, ** at 5 percent level, * at 10 percent level.
}
\end{table}

\begin{table}[htb]
\begin{center}
\caption{The effect of treatment on subsequent number of users  and edits, pages in German Wikipedia}
\label{T:RegrRobustUsers_de}
\renewcommand{\tabcolsep}{9pt}
\begin{tabular}{ l cc cc }
\hline
  & 1st year & 2nd year & 3rd year & 4th year \\  & (1) & (2) & (3) & (4) \\ \hline \multicolumn{5}{c}{Panel A. Dependent variable: average number of users per month} \\ 
Treatment group     &       0.219***&       0.131   &       0.097   &       0.064   \\
                    &     (0.082)   &     (0.080)   &     (0.068)   &     (0.072)   \\
Mean dep. var.      &       0.510   &       0.357   &       0.360   &       0.354   \\
SD dep. var.        &       0.332   &       0.314   &       0.264   &       0.280   \\
Adj. R-squared      &       0.096   &       0.028   &       0.018   &      -0.004   \\
Observations        &          60   &          60   &          60   &          60   \\
 \hline  \multicolumn{5}{c}{Panel B. Dependent variable: average number of edits per month} \\ 
Treatment group     &       0.239** &       0.139   &       0.108   &       0.083   \\
                    &     (0.092)   &     (0.084)   &     (0.077)   &     (0.080)   \\
Mean dep. var.      &       0.533   &       0.372   &       0.396   &       0.375   \\
SD dep. var.        &       0.373   &       0.329   &       0.299   &       0.308   \\
Adj. R-squared      &       0.089   &       0.029   &       0.017   &       0.002   \\
Observations        &          60   &          60   &          60   &          60   \\

\hline
\end{tabular}
\end{center}
\footnotesize{
Notes: Each column presents estimates from a separate cross-section regression of 60 Wikipedia pages in German Wikipedia. 
Dependent variable is the average number of users (panel A) or edits (panel B) per month during the 1st year post-treatment (column 1), 2nd (column 2), 3rd (column 3), and the 4th year post-treatment (column 4). 
Standard errors are reported in parentheses. *** Indicates significance at the 1 percent level, ** at 5 percent level, * at 10 percent level.
}
\end{table}

\begin{table}[htb]
\begin{center}
\caption{The effect of treatment on subsequent number of users and edits, pages in Italian Wikipedia}
\label{T:RegrRobustUsers_it}
\renewcommand{\tabcolsep}{9pt}
\begin{tabular}{ l cc cc }
\hline
  & 1st year & 2nd year & 3rd year & 4th year \\  & (1) & (2) & (3) & (4) \\ \hline \multicolumn{5}{c}{Panel A. Dependent variable: average number of users per month} \\ 
Treatment group     &      -0.061   &      -0.025   &      -0.097   &      -0.064   \\
                    &     (0.053)   &     (0.069)   &     (0.066)   &     (0.063)   \\
Mean dep. var.      &       0.203   &       0.221   &       0.201   &       0.196   \\
SD dep. var.        &       0.205   &       0.265   &       0.258   &       0.243   \\
Adj. R-squared      &       0.006   &      -0.015   &       0.020   &       0.001   \\
Observations        &          60   &          60   &          60   &          60   \\
 \hline  \multicolumn{5}{c}{Panel B. Dependent variable: average number of edits per month} \\ 
Treatment group     &      -0.058   &      -0.028   &      -0.106   &      -0.069   \\
                    &     (0.054)   &     (0.072)   &     (0.070)   &     (0.068)   \\
Mean dep. var.      &       0.204   &       0.225   &       0.208   &       0.207   \\
SD dep. var.        &       0.209   &       0.277   &       0.273   &       0.264   \\
Adj. R-squared      &       0.003   &      -0.015   &       0.021   &       0.001   \\
Observations        &          60   &          60   &          60   &          60   \\

\hline
\end{tabular}
\end{center}
\footnotesize{
Notes: Each column presents estimates from a separate cross-section regression of 60 Wikipedia pages in Italian Wikipedia. 
Dependent variable is the average number of users (panel A) or edits (panel B) per month during the 1st year post-treatment (column 1), 2nd (column 2), 3rd (column 3), and the 4th year post-treatment (column 4). 
Standard errors are reported in parentheses. *** Indicates significance at the 1 percent level, ** at 5 percent level, * at 10 percent level.
}
\end{table}

\clearpage 

\begin{table}[htb]
\begin{center}
\caption{The effect of treatment on subsequent number of edits by old versus new users}
\label{T:RegrRobustEditsByOldUsersCS}
\renewcommand{\tabcolsep}{3pt}
\begin{tabular}{ l cc cc cc cc }
\hline
 & \multicolumn{2}{c}{1st year} & \multicolumn{2}{c}{2nd year} & \multicolumn{2}{c}{3rd year} & \multicolumn{2}{c}{4th year} \\  & (1) & (2) & (3) & (4) & (5) & (6) & (7) & (8)\\ \hline \multicolumn{9}{c}{Panel A. Dependent variable: Average number of edits by old users per month} \\ 
Treatment group     &      -0.004   &       0.006   &      -0.010   &      -0.011   &      -0.006   &      -0.007   &       0.004   &       0.008   \\
                    &     (0.016)   &     (0.015)   &     (0.009)   &     (0.009)   &     (0.012)   &     (0.012)   &     (0.015)   &     (0.015)   \\
Language FE         &          No   &         Yes   &          No   &         Yes   &          No   &         Yes   &          No   &         Yes   \\
City FE             &          No   &         Yes   &          No   &         Yes   &          No   &         Yes   &          No   &         Yes   \\
Mean dep. var.      &       0.070   &       0.070   &       0.037   &       0.037   &       0.041   &       0.041   &       0.041   &       0.041   \\
SD dep. var.        &       0.107   &       0.107   &       0.061   &       0.061   &       0.080   &       0.080   &       0.103   &       0.103   \\
Adj. R-squared      &      -0.005   &       0.201   &       0.001   &       0.042   &      -0.004   &       0.044   &      -0.005   &       0.147   \\
Observations        &         180   &         180   &         180   &         180   &         180   &         180   &         180   &         180   \\
 \hline  \multicolumn{9}{c}{Panel B. Dependent variable: Average number of edits by new users per month} \\ 
Treatment group     &       0.113** &       0.133***&       0.129** &       0.151***&      -0.002   &       0.018   &      -0.000   &       0.022   \\
                    &     (0.044)   &     (0.029)   &     (0.057)   &     (0.040)   &     (0.045)   &     (0.027)   &     (0.038)   &     (0.031)   \\
Language FE         &          No   &         Yes   &          No   &         Yes   &          No   &         Yes   &          No   &         Yes   \\
City FE             &          No   &         Yes   &          No   &         Yes   &          No   &         Yes   &          No   &         Yes   \\
Mean dep. var.      &       0.319   &       0.319   &       0.299   &       0.299   &       0.278   &       0.278   &       0.264   &       0.264   \\
SD dep. var.        &       0.301   &       0.301   &       0.387   &       0.387   &       0.298   &       0.298   &       0.255   &       0.255   \\
Adj. R-squared      &       0.030   &       0.637   &       0.022   &       0.568   &      -0.006   &       0.662   &      -0.006   &       0.416   \\
Observations        &         180   &         180   &         180   &         180   &         180   &         180   &         180   &         180   \\

\hline
\end{tabular}
\end{center}
\footnotesize{
Notes: 
Each column presents estimates from a separate cross-section regression of 180 Wikipedia pages. Dependent variable is the average number of edits by \textit{old users} (panel A) or \textit{new users} (panel B) per month during the 1st year post-treatment (columns 1--2), 2nd (columns 3--4), 3rd (columns 5--6), and the 4th year post-treatment (columns 7--8). \textit{Old users} are defined as those who had edited the page before treatment, and \textit{New users} are those who had not. 
Standard errors are reported in parentheses. *** Indicates significance at 1 percent level, ** at 5 percent level, * at 10 percent level. 
}
\end{table}

\begin{table}[htb]
\begin{center}
\caption{The effect of treatment on subsequent number of users per month, by page age}
\label{T:RegrHeterUsersPageAge}
\renewcommand{\tabcolsep}{3pt}
\begin{tabular}{ l cc cc cc cc }
\hline
  & \multicolumn{2}{c}{1st year} & \multicolumn{2}{c}{2nd year} & \multicolumn{2}{c}{3rd year} & \multicolumn{2}{c}{4th year} \\  & (1) & (2) & (3) & (4) & (5) & (6) & (7) & (8) \\ \hline & \multicolumn{8}{c}{Panel A. Sample: pages with age below median} \\ 
Treatment group     &       0.172***&       0.176***&       0.103   &       0.134*  &       0.042   &       0.034   &      -0.033   &      -0.032   \\
                    &     (0.054)   &     (0.055)   &     (0.070)   &     (0.076)   &     (0.053)   &     (0.055)   &     (0.054)   &     (0.054)   \\
Language FE         &          No   &         Yes   &          No   &         Yes   &          No   &         Yes   &          No   &         Yes   \\
Mean dep. var.      &       0.303   &       0.303   &       0.232   &       0.232   &       0.246   &       0.246   &       0.217   &       0.217   \\
SD dep. var.        &       0.224   &       0.224   &       0.275   &       0.275   &       0.206   &       0.206   &       0.207   &       0.207   \\
Adj. R-squared      &       0.136   &       0.134   &       0.019   &       0.033   &      -0.007   &      -0.001   &      -0.011   &      -0.020   \\
Observations        &          60   &          60   &          60   &          60   &          60   &          60   &          60   &          60   \\
 \hline  & \multicolumn{8}{c}{Panel B. Sample: pages with age above median} \\ 
Treatment group     &       0.062   &       0.072   &       0.093   &       0.106   &      -0.029   &      -0.035   &       0.015   &       0.019   \\
                    &     (0.067)   &     (0.060)   &     (0.069)   &     (0.075)   &     (0.058)   &     (0.060)   &     (0.057)   &     (0.055)   \\
Language FE         &          No   &         Yes   &          No   &         Yes   &          No   &         Yes   &          No   &         Yes   \\
Mean dep. var.      &       0.401   &       0.401   &       0.359   &       0.359   &       0.320   &       0.320   &       0.330   &       0.330   \\
SD dep. var.        &       0.367   &       0.367   &       0.381   &       0.381   &       0.318   &       0.318   &       0.310   &       0.310   \\
Adj. R-squared      &      -0.001   &       0.306   &       0.007   &       0.178   &      -0.006   &       0.188   &      -0.008   &       0.187   \\
Observations        &         120   &         120   &         120   &         120   &         120   &         120   &         120   &         120   \\

\hline
\end{tabular}
\end{center}
\footnotesize{
Notes: Each column presents estimates from a separate cross-section regression of Wikipedia pages. Dependent variable is the average number of users per month during the 1st year post-treatment (columns 1--2), 2nd (columns 3--4), 3rd (columns 5--6), and the 4th year post-treatment (columns 7--8). Sample is restricted to pages with age below median (panel A) or above median (panel B). The sample is split to unequal groups because many pages were created on a same day.
Standard errors are reported in parentheses. *** Indicates significance at 1 percent level, ** at 5 percent level, * at 10 percent level.
}
\end{table}

\clearpage 

\begin{table}[h]
\begin{center}
\caption{Robustness: the effect of treatment on subsequent edit distance, alternative controls}
\label{T:RegrRobustEditDistGroupFE} 
\renewcommand{\tabcolsep}{11pt}
\begin{tabular}{ l cc cc } 
\hline
 & 1st year & 2nd year & 3rd year & 4th year \\  & (1) & (2) & (3) & (4) \\ \hline  \multicolumn{5}{c}{Panel A. Dependent variable: monthly average edit distance} \\ 
Treatment group     &       7.655   &     128.524   &     -15.897   &      23.890   \\
                    &    (30.583)   &   (101.061)   &    (21.408)   &    (73.634)   \\
Group FE            &         Yes   &         Yes   &         Yes   &         Yes   \\
Mean dep. var.      &     119.848   &     108.318   &      65.123   &     112.119   \\
SD dep. var.        &     205.394   &     678.724   &     145.853   &     500.082   \\
Adj. R-squared      &       0.002   &       0.002   &       0.031   &       0.024   \\
Observations        &         180   &         180   &         180   &         180   \\
 \hline \multicolumn{5}{c}{Panel A. Dependent variable: monthly average capped edit distance} \\ 
Treatment group     &      13.644** &      18.642*  &      -6.174   &      -2.823   \\
                    &     (6.116)   &     (9.786)   &     (6.699)   &     (6.004)   \\
Group FE            &         Yes   &         Yes   &         Yes   &         Yes   \\
Mean dep. var.      &      39.173   &      35.008   &      32.183   &      28.758   \\
SD dep. var.        &      41.530   &      66.061   &      45.675   &      41.095   \\
Adj. R-squared      &       0.024   &       0.012   &       0.032   &       0.039   \\
Observations        &         180   &         180   &         180   &         180   \\

\hline
\end{tabular}
\end{center}
\footnotesize{
Notes: Each column presents estimates from a separate cross-section regression of 180 Wikipedia pages. Edit distance (panel A) equals the number of characters added plus the number of characters deleted. Capped edit distance (panel B) is calculated using individual edits which edit distance is capped at the 90th percentile. All dependent variables are averages across months and are measured in the number of characters. All regressions include stratification group dummies, \textit{Group FE}.  Standard errors are reported in parentheses. *** Indicates significance at 1 percent level, ** at 5 percent level, * at 10 percent level.
} 
\end{table}

\begin{table}[h]
\begin{center}
\caption{Robustness: the effect of treatment on subsequent edit distance, pre-treatment average edit distance as a control}
\label{T:RegrRobustEditDistControlPastVar} 
\renewcommand{\tabcolsep}{0pt}
\begin{tabular}{ l cc cc cc cc } 
\hline
 & \multicolumn{2}{c}{1st year} & \multicolumn{2}{c}{2nd year} & \multicolumn{2}{c}{3rd year} & \multicolumn{2}{c}{4th year} \\  & (1) & (2) & (3) & (4) & (5) & (6) & (7) & (8) \\ \hline  & \multicolumn{8}{c}{Panel A. Dependent variable: monthly average edit distance} \\ 
Treatment group     &       7.929   &      14.054   &     130.293   &     103.817   &     -14.643   &     -14.895   &      25.234   &      -0.361   \\
                    &    (30.731)   &    (31.036)   &   (100.588)   &   (105.085)   &    (20.124)   &    (21.084)   &    (74.399)   &    (79.345)   \\
Aver. edit dist.    &       0.123   &      -0.369*  &       0.794   &      -1.043   &       0.563***&       0.402***&       0.603   &       0.489   \\
before treatment    &     (0.154)   &     (0.210)   &     (0.503)   &     (0.712)   &     (0.101)   &     (0.143)   &     (0.372)   &     (0.538)   \\
Language FE         &          No   &         Yes   &          No   &         Yes   &          No   &         Yes   &          No   &         Yes   \\
City FE             &          No   &         Yes   &          No   &         Yes   &          No   &         Yes   &          No   &         Yes   \\
Mean dep. var.      &     119.848   &     119.848   &     108.318   &     108.318   &      65.123   &      65.123   &     112.119   &     112.119   \\
SD dep. var.        &     205.394   &     205.394   &     678.724   &     678.724   &     145.853   &     145.853   &     500.082   &     500.082   \\
Adj. R-squared      &      -0.007   &       0.087   &       0.012   &       0.042   &       0.143   &       0.165   &       0.004   &      -0.006   \\
Observations        &         180   &         180   &         180   &         180   &         180   &         180   &         180   &         180   \\
 \hline & \multicolumn{8}{c}{Panel A. Dependent variable: monthly average capped edit distance} \\ 
Treatment group     &      15.554***&      17.414***&      22.063** &      21.242** &      -3.341   &      -3.601   &      -0.533   &       1.048   \\
                    &     (5.549)   &     (5.629)   &     (8.594)   &     (8.527)   &     (5.599)   &     (5.158)   &     (5.284)   &     (5.204)   \\
Aver. capped edit   &       0.447***&       0.072   &       0.801***&       0.394** &       0.663***&       0.300***&       0.536***&       0.304***\\
dist. before treatment&     (0.070)   &     (0.120)   &     (0.109)   &     (0.182)   &     (0.071)   &     (0.110)   &     (0.067)   &     (0.111)   \\
Language FE         &          No   &         Yes   &          No   &         Yes   &          No   &         Yes   &          No   &         Yes   \\
City FE             &          No   &         Yes   &          No   &         Yes   &          No   &         Yes   &          No   &         Yes   \\
Mean dep. var.      &      39.173   &      39.173   &      35.008   &      35.008   &      32.183   &      32.183   &      28.758   &      28.758   \\
SD dep. var.        &      41.530   &      41.530   &      66.061   &      66.061   &      45.675   &      45.675   &      41.095   &      41.095   \\
Adj. R-squared      &       0.199   &       0.266   &       0.241   &       0.335   &       0.326   &       0.491   &       0.258   &       0.360   \\
Observations        &         180   &         180   &         180   &         180   &         180   &         180   &         180   &         180   \\

\hline
\end{tabular}
\end{center}
\footnotesize{
Notes: Each column presents estimates from a separate cross-section regression of 180 Wikipedia pages. Edit distance (panel A) equals the number of characters added plus the number of characters deleted. Capped edit distance (panel B) is calculated using individual edits which edit distance is capped at the 90th percentile. All dependent variables are averages across months and are measured in the number of characters. All regressions include as a covariate pre-treament  average edit distance or capped edit distance. Standard errors are reported in parentheses. *** Indicates significance at 1 percent level, ** at 5 percent level, * at 10 percent level.
} 
\end{table}

\begin{table}[h!]
\begin{center}
\caption{Robustness: The effect of treatment on subsequent  edit distance measures, panel data FE regression}
\label{T:RegrRobustEditDistPanel} 
\renewcommand{\tabcolsep}{6pt}
\begin{tabular}{ l cccc} 
\hline
 & \multicolumn{2}{c}{Edit distance} & \multicolumn{2}{c}{Capped edit} \\ & \multicolumn{2}{c}{} & \multicolumn{2}{c}{distance} \\ & (1) & (2) & (3) & (4)  \\ \hline 
Treatment group,    &       9.882   &       9.882   &      17.917***&      17.917***\\
post-treatment 1st year&    (33.347)   &    (33.005)   &     (6.419)   &     (6.413)   \\
Treatment group,    &     130.751   &     130.751   &      22.914***&      22.914***\\
post-treatment 2nd year&   (100.353)   &   (100.515)   &     (8.638)   &     (8.580)   \\
Treatment group,    &     -13.670   &     -13.670   &      -1.901   &      -1.901   \\
post-treatment 3rd year&    (21.110)   &    (21.168)   &     (5.920)   &     (5.918)   \\
Treatment group,    &      26.117   &      26.117   &       1.449   &       1.449   \\
post-treatment 4th year&    (74.428)   &    (74.629)   &     (5.932)   &     (5.809)   \\
Month FE            &         Yes   &         Yes   &         Yes   &         Yes   \\
Month-language FE   &          No   &         Yes   &          No   &         Yes   \\
Page FE             &         Yes   &         Yes   &         Yes   &         Yes   \\
Mean dep. var.      &      88.936   &      88.936   &      37.271   &      37.271   \\
SD dep. var.        &    1086.255   &    1086.255   &     134.352   &     134.352   \\
Observations        &       18360   &       18360   &       18360   &       18360   \\

\hline
\end{tabular}
\end{center}
\footnotesize{
Notes: A unit of observation is a page-month pair. 
Dependent variable is monthly edit distance (columns 1-2) or capped edit distance (columns 3-4). Edit distance equals the number of characters added plus the number of characters deleted. Capped edit distance is calculated using individual edits which edit distance is capped at the 90th percentile. All dependent variables are averages across months and are measured in the number of characters. 
All regressions include page fixed effects and either month fixed effects or month-language fixed effects. 
\textit{Treatment group, post-treatment 1st year} is an indicator variable that takes value one during the first year post-treatment if the page belongs to the treatment group and zero otherwise;  and similarly for other years. 
The sample is a balanced sample that starts from the time period when all the 180 pages existed. 
Standard errors, reported in parentheses, are clustered by page (180 pages). *** Indicates significance at the 1 percent level, ** at 5 percent level, * at 10 percent level.
} 
\end{table}


\end{document}